\documentclass[12pt,dvips]{article}
\usepackage[dvips]{graphics}

\renewcommand{\theequation}{\thesection.\arabic{equation}}

\newlength{\minitwocolumn}
\def\beq{\begin{equation}}
\def\eeq{\end{equation}}
\def\bea{\begin{eqnarray}}
\def\eea{\end{eqnarray}}
\def\bseq{\begin{subequations}}
\def\eseq{\end{subequations}}
\def\nn{\nonumber}
\def\dfrac{\displaystyle\frac}
\def\numt#1#2{#1 \times 10^{#2}}
\def\etal{{\it et al.}}

\def\ie{{\it i.e.,~}}

\def\bs{\bigskip}

\def\btiny{\begin{tiny}}
\def\etiny{\end{tiny}}
\def\bsc{\begin{scriptsize}}
\def\esc{\end{scriptsize}}
\def\bfoot{\begin{footnotesize}}
\def\efoot{\end{footnotesize}}
\def\bsm{\begin{small}}
\def\esm{\end{small}}
\def\bno{\begin{normalsize}}
\def\eno{\end{normalsize}}
\def\bla{\begin{large}}
\def\ela{\end{large}}
\def\bLa{\begin{Large}}
\def\eLa{\end{Large}}
\def\bLA{\begin{LARGE}}
\def\eLA{\end{LARGE}}
\def\bhu{\begin{huge}}
\def\ehu{\end{huge}}
\def\bHu{\begin{Huge}}
\def\eHu{\end{Huge}}

\def\bCe{\begin{center}}
\def\eCe{\end{center}}
\def\bFR{\begin{flushright}}
\def\eFR{\end{flushright}}
\def\bFL{\begin{flushleft}}
\def\eFL{\end{flushleft}}

\def\UL{\underline}

\def\PR#1#2#3{Phys. Rev. {\bf #1}, #2 (#3)}
\def\PRL#1#2#3{Phys. Pev. Lett. {\bf #1}, #2 (#3)}
\def\PL#1#2#3{Phys. Lett. {\bf #1}, #2 (#3)}

\def\NP#1#2#3{Nucl. Phys. {\bf #1}, #2 (#3)}

\def\PTP#1#2#3{Prog. Theor. Phys. {\bf #1}, #2 (#3)}

\def\eqref#1{eq.(\ref{eqn:#1})}
\def\Eqref#1{Equation~(\ref{eqn:#1})}
\def\eqsref#1{eqs.(\ref{eqn:#1})}
\def\Eqsref#1{Equations~(\ref{eqn:#1})}
\def\eqvref#1{(\ref{eqn:#1})}

\def\eqlab#1{\label{eqn:#1}}

\def\Tbref#1{Table~\ref{tbl:#1}}

\def\tblab#1{\label{tbl:#1}}

\def\Fgref#1{Figure~\ref{fig:#1}}

\def\Fgsref#1{Figures~\ref{fig:#1}}

\def\Fgvref#1{\ref{fig:#1}}
\def\fglab#1{\label{fig:#1}}

\def\bmaT{\left(\begin{array}{ccc}}
\def\emaT{\end{array}\right)}
\def\bma{\left( \begin{array} }
\def\ema{\end{array} \right)}

\def\vev#1{\langle #1 \rangle}

\def\l{\left}
\def\r{\right}

\def\Su{\rm u}
\def\Sd{\rm d}

\def\Se{\rm e}

\def\tr{{\rm tr}}

\def\hu{{y^{\rm u}}}

\def\he{{y^{\rm e}}}

\def\s23#1{{ \sin {#1} {\theta_{23}}}}
\def\c23#1{{ \cos {#1} {\theta_{23}}}}

\makeatletter
\newtoks\@stequation

\def\subequations{\refstepcounter{equation}%
\edef\@savedequation{\the\c@equation}%
\@stequation=\expandafter{\theequation}
\edef\@savedtheequation{\the\@stequation}
\edef\oldtheequation{\theequation}%
\setcounter{equation}{0}%
\def\theequation{\oldtheequation\alph{equation}}}

\def\endsubequations{%
\ifnum\c@equation < 2 \@warning{Only \the\c@equation\space subequation
used in equation \@savedequation}\fi
\setcounter{equation}{\@savedequation}%
\@stequation=\expandafter{\@savedtheequation}%
\edef\theequation{\the\@stequation}%
\global\@ignoretrue}

\def\eqnarray{\stepcounter{equation}\let\@currentlabel\theequation
\global\@eqnswtrue\m@th
\global\@eqcnt\z@\tabskip\@centering\let\\\@eqncr
$$\halign to\displaywidth\bgroup\@eqnsel\hskip\@centering
$\displaystyle\tabskip\z@{##}$&\global\@eqcnt\@ne
\hfil$\;{##}\;$\hfil
&\global\@eqcnt\tw@ $\displaystyle\tabskip\z@{##}$\hfil
\tabskip\@centering&\llap{##}\tabskip\z@\cr}

\makeatother
\begin{document}

\title{
Stability of the Lepton-Flavor Mixing Matrix\\
 Against Quantum Corrections
}
\author{
{N. Haba$^{1,2}$}\thanks{E-mail address:haba@pacific.mps.ohio-state.edu}
\quad
and
\quad
{N. Okamura$^3$}\thanks{E-mail address:naotoshi.okamura@kek.jp}\\
\\
{\small \it $^1$Department of Physics, The Ohio State University,}
{\small \it Columbus, Ohio 43210, USA}\\
{\small \it $^2$Faculty of Engineering, Mie University,}
{\small \it Tsu Mie 514-8507, Japan}\\
{\small \it $^3$Theory Group, KEK, Tsukuba Ibaraki 305-0801, Japan}
}
\date{}
\maketitle
\vspace{-10.5cm}
\begin{flushright}
hep-ph/9906481 \\
OHSTPY-HEP-T-99-013 \\
KEK-TH-632 
\end{flushright}
\vspace{10.5cm}
\vspace{-2.5cm}
\begin{abstract}
 Recent neutrino experiments suggest the strong evidences of 
tiny neutrino masses and the lepton-flavor mixing. 
 Neutrino-oscillation solutions for
the atmospheric neutrino anomaly and the solar neutrino deficit
can determine the texture of neutrino mass matrix
according to the neutrino mass hierarchies as
 Type A: $m_3^{} \gg  m_2^{} \sim m_1^{}$,
 Type B: $m_3^{} \ll  m_2^{} \sim m_1^{}$, and
 Type C: $m_3^{} \sim m_2^{} \sim m_1^{}$.
In this paper we study the stability
 of the lepton-flavor mixing matrix against quantum corrections
for all types of mass hierarchy
 in the minimal supersymmetric Standard Model
 with the effective dimension-five operator which
 gives Majorana masses of neutrinos.
The relative sign assignments of neutrino masses in each type
 play the crucial roles for the stability against
quantum corrections.
 We find the lepton-flavor mixing matrix of 
Type A is stable against quantum corrections, and
that of Type B with the same (opposite)
signs of $m_1^{}$ and $m_2^{}$ are unstable (stable).
 For Type C, the lepton-flavor-mixing
matrix approaches to the definite unitary matrix according to the
relative sign assignments of neutrino mass eigenvalues,
as the effects of quantum corrections become
large enough to neglect squared mass differences of neutrinos.
\end{abstract}

{\sf PACS NUMBER:\\
\hspace*{5ex}12.15.Ff, 14.60.Pq, 23.40.Bw }

{\sf KEY WORDS:\\
\hspace*{5ex} Lepton Flavor Mixing, Renormalization Effect}

\newpage
\section{Introduction}
\label{sec:1}
\setcounter{equation}{0}
 
Recent neutrino experiments suggest the strong evidences of 
 tiny neutrino masses 
and  the flavor mixing in the lepton sector
\cite{solar4}\nocite{Atm4,SK4}-\cite{CHOOZ}.
Studies of the lepton-flavor-mixing matrix,
 which is called Maki-Nakagawa-Sakata (MNS) matrix \cite{MNS},
 point to new steps of the flavor physics.
Especially, study of the 
energy-scale dependence of the MNS matrix
 is one of the main issues to investigate the new physics 
 beyond the Standard Model (SM) \cite{up2now}.

There are following neutrino-oscillation solutions for
the solar neutrino deficit and the atmospheric neutrino anomaly:
\bseq
\bea
\Delta m_{\rm solar}^2 &=&
\l\{
\begin{array}{cl}
0.85 \times 10^{-10}& \mbox{eV$^2$~~(vacuum solution), } \\
1.8  \times 10^{-5} & \mbox{eV$^2$~~(MSW-large mixing solution),}  \\
0.8  \times 10^{-5} & \mbox{eV$^2$~~(MSW-small mixing solution),}  
\end{array}
\r. \eqlab{mass-SUN} \\
\nn \\
\Delta m_{\rm ATM}^2 &=&3.7  \times 10^{-3} \mbox{~~eV$^2$}\,,
\eqlab{mass-ATM}
\eea
\eqlab{mass-h2}
\eseq

\vspace*{-3ex}

\noindent
where $\Delta m_{\rm solar}^2$ and $\Delta m_{\rm ATM}^2$ 
stand for the squared mass differences 
of the solar neutrino deficit \cite{solar4}
and the atmospheric neutrino anomaly
\cite{Atm4,SK4}, respectively.
 In this article, we adopt the scenario of three generation neutrinos
which means
\beq
\Delta m_{\rm solar}^2 \equiv \l|m_2^2-m_1^2\r|\,,
\mbox{\quad and \quad}
\Delta m_{\rm ATM}^2 \equiv \l|m_3^2-m_2^2\r|\,,
\eqlab{mass-h}
\eeq
where $m_i$ is the $i$-th ($i=1\sim 3$)
generation neutrino mass ($m_i \geq 0$).
We take the following values of the mixing angles
for their solutions.
\bseq
\bea
\sin^2 2\theta_{\rm solar}
&=&
\l\{
\begin{array}{ll}
1    & \mbox{~~(vacuum solution), } \\
1    & \mbox{~~(MSW-large mixing solution),}  \\
0.017& \mbox{~~(MSW-small mixing solution),}  
\end{array}
\r. \eqlab{angle-SUN} \\
\sin^2 2\theta_{\rm ATM}
 &=& 1 \mbox{~~~~~~~~~~~~~(atmospheric neutrino anomaly)}\,,
\eqlab{angle-ATM} \\
\sin^2 2\theta_{\rm CHOOZ} &=& 0 \mbox{~~~~~~~~~~~~~(CHOOZ experiment)}\,.
\eqlab{angle-chooz}
\eea
\eqlab{angle}
\eseq

\noindent

Under the assignments of \eqsref{mass-h},
there are three possible types of neutrino
mass hierarchy \cite{Altarelli} as
\bseq
\bea
\mbox{Type A}
&~:~& 
m_3 \gg m_2 \sim m_1 \,,
\eqlab{typeA} \\
\mbox{Type B}
&~:~& m_1 \sim m_2 \gg m_3\,,
\eqlab{typeB}\\
\mbox{Type C}
&~:~& m_3 \sim m_2 \sim m_1\,.
\eqlab{typeC}
\eea
\eqlab{type}
\eseq

\vspace*{-3ex}

\noindent

In this article,
 we study the stability of the MNS matrix
 for these three types of neutrino mass hierarchy
 against quantum corrections
 in the minimal supersymmetric Standard Model (MSSM)
 with the effective dimension-five operator which
 gives Majorana masses of neutrinos.
 Since the negative sign assignments 
of $m_i$ in \eqsref{type}
also satisfy \eqsref{mass-h},
 we consider all relative sign 
assignments for masses in each type.
 We determine
the MNS matrix elements at the low energy scale
from results of neutrino-oscillation experiments,
and analyze whether the MNS matrix
is stable against quantum corrections or not
in each type of neutrino mass hierarchy.
The results of analyses strongly depend on
the types of neutrino mass hierarchy and
the relative sign assignments of masses
as follows.

\bs

\noindent
\UL{Type A}:
The MNS matrix is stable against quantum corrections.

\noindent
\UL{Type B}: 
The MNS matrix is unstable (stable)
when $m_1^{}$ and $m_2^{}$ have the same (opposite) signs.

\noindent
\UL{Type C}:
The MNS matrix approaches to the definite unitary matrix
according to the relative sign assignments of neutrino masses,
as the effects of quantum corrections become
large enough to neglect squared masse differences
of neutrino.

\bs

 We also study the stability of the MNS matrix
against quantum corrections in the two generation neutrinos.
 Results of this article are not only
useful for the model building beyond the SM,
but also show the possibility to obtain the large mixing angles
from quantum corrections.

 This article is organized as follows.
 In section \ref{sec:MNS}, 
 we determine the elements of 
the MNS matrix from the data of recent neutrino-oscillation
experiments.
 In section \ref{sec:RGE}, 
 we estimate the magnitude of quantum corrections
of the dimension-five Majorana operator.
 In section \ref{sec:Toy}, 
 we study the stability of the MNS matrix
against renormalization effects in the two generation
neutrinos.
 In section \ref{sec:3gen2}, 
 we analyze the stability of the MNS matrix against
quantum corrections in the three generation neutrinos
for each type of mass hierarchy.
 Section \ref{sec:LAST} gives the conclusion.

\section{The Maki-Nakagawa-Sakata Matrix}
\label{sec:MNS}
\setcounter{equation}{0}
 In this section, we give the definition and the
parameterization of the Maki-Nakagawa-Sakata (MNS) 
lepton-flavor mixing matrix \cite{MNS}.
 We determine elements of 
the MNS matrix from the data of recent neutrino-oscillation
experiments \cite{solar4}\nocite{Atm4,SK4}-\cite{CHOOZ}. 

\subsection{Definition}
 The effective Yukawa Lagrangian in the lepton sector is given by
\beq
{\cal
L}^{\rm low}_{yukawa}
           =  {\he}_{ij} \phi_{\Sd} L_i \cdot e_{Rj}^c
            - \dfrac{1}{2} \kappa_{ij}(\phi_{\Su}L_i)\cdot(\phi_{\Su}L_j)
            + h.c.\,,
\eqlab{W_low}
\eeq
where ${y^{\rm e}_{ij}}$ is
the Yukawa matrix of the charged lepton.
$\phi_{\Su}$ and $\phi_{\Sd}$
are the SU(2)$_L$ doublet Higgs bosons
that give Dirac masses to the up-type and the down-type 
fermions, respectively.
 $L_i$ is the $i$-th generation SU(2)$_L$ 
doublet lepton.
 $e_{Ri}^{}$ is the $i$-th generation charged lepton.
 $\kappa$ induces the neutrino Majorana mass matrix
which is complex and symmetric.

 Now we give the definition of the $3 \times 3$ MNS matrix,
which is defined on the analogy of the 
CKM matrix \cite{CKM,PDG}.
Unitary matrices of $U_{\Se}$ and $U_{\nu}$
transform the mass-eigenstates into the
weak-current eigenstates as
\beq
\bma{c}
l_{L1}^{} \\
l_{L2}^{} \\
l_{L2}^{}
\ema
 =
U_{\Se}
\bma{c}
e_{L}^{} \\
\mu_{L}^{} \\
\tau_{L}^{}
\ema
\,,~~~~~~~~~~
\bma{c}
\nu_{L1}^{} \\
\nu_{L2}^{} \\
\nu_{L3}^{}
\ema
 =
U_{\nu}
\bma{c}
\nu_1^{} \\
\nu_2^{} \\
\nu_3^{}
\ema
\,.
\eqlab{UeUn}
\eeq
 The MNS matrix is then defined as
\beq
\eqlab{def_MNS}
\l(V_{\rm MNS}^{}\r)_{\alpha i} \equiv
\l( U_{\Se}^{\dagger} U_{\nu}^{} \r)_{\alpha i}^{}\,,
\mbox{\quad \quad}
\nu_\alpha^{}=\sum^3_{i=1} \l(V_{\rm MNS}^{}\r)_{\alpha i}^{} \nu_i^{}\,,
\eeq
where $\alpha$ and $i$ label the neutrino flavors
($\alpha=e, \mu, \tau$)
and the mass eigenstates ($i=1,2,3$), respectively.

\subsection{Parameterization}
The $3\times3$ MNS matrix has three mixing angles
and three physical phases in general for the Majorana neutrinos.
We adopt the parameterization of
\beq
V_{\rm MNS}^{} =
\underbrace{
\bmaT
U_{e 1}    & U_{e 2}    & U_{e 3} \\
U_{\mu 1}  & U_{\mu 2}  & U_{\mu 3} \\
U_{\tau 1} & U_{\tau 2} & U_{\tau 3}  
\emaT
}_{U_{\alpha i}}
\bmaT
1 & 0 & 0 \\
0 & e^{i \varphi_2^{}} & 0 \\
0 & 0 & e^{i \varphi_3^{}} 
\emaT\,.
\eqlab{MNS1}
\eeq
 The matrix $U_{\alpha i}$,
which has three mixing angles and one phase,
can be parameterized as the same way as the CKM matrix.
 Since the data of 
the present neutrino-oscillation experiments directly 
constrain elements of $U_{e2}$, $U_{e3}$, and $U_{\mu 3}$,
the most convenient parameterization is 
to adopt these three elements
as the independent parameters \cite{Wolf,MK,no6}.
 Without losing generality, we can take $U_{e2}$
and $U_{\mu 3}$ to be real and non-negative by the
redefinition of $\varphi_2^{}$ and $\varphi_3^{}$.
 Then only $U_{e3}$ has a complex phase.
 All the other matrix elements can be determined by the
unitarity conditions as follows.

\begin{minipage}[t]{\minitwocolumn}
\bea
U_{e1} &=& \sqrt{1-|U_{e3}|^2-|U_{e2}|^2}\,, \nn \\
\nn \\
U_{\mu 1} &=& - \dfrac{U_{e2}U_{\tau 3} + U_{\mu 3}U_{e1}U_{e3}^{\ast} }
                      {1-|U_{e3}|^2}\,,  \nn \\
\nn \\
U_{\tau 1}&=& \dfrac{U_{e2}U_{\mu 3} - U_{\tau 3}U_{e1}U_{e3}^{\ast} }
                     {1-|U_{e3}|^2}\,, \nn 
\eea
\end{minipage}
\hspace{\columnsep}
\begin{minipage}[t]{\minitwocolumn}
\bea
U_{\tau 3} &=& \sqrt{1-|U_{e3}|^2-|U_{\mu 3}|^2}\,,\nn \\ 
\nn \\
U_{\mu 2}&=& \dfrac{U_{e1}U_{\tau 3} - U_{\mu 3}U_{e2}U_{e3}^{\ast} }
                    {1-|U_{e3}|^2}\,, \mbox{~~~~~~~~~}
\eqlab{MNS-RC}\\
\nn \\
U_{\tau 2} &=& - \dfrac{U_{\mu 3}U_{e1} + U_{e2}U_{\tau 3}U_{e3}^{\ast} }
                       {1-|U_{e3}|^2}\,.\nn
\eea
\end{minipage}

\noindent
Here $U_{e1}$, $U_{e2}$, $U_{\mu 3}$, and $U_{\tau 3}$
are real and non-negative,
and the other elements are complex.

 The relations among the MNS matrix elements
and the mixing angles are given by
\beq
\sin \theta_{13} = |U_{e3}|\,,
\mbox{\quad}
\sin \theta_{23} = \dfrac{U_{\mu 3}}{\sqrt{1-|U_{e3}|^2}}\,,
\mbox{\quad}
\sin \theta_{12} = \dfrac{U_{e 2}}{\sqrt{1-|U_{e3}|^2}}\,,
\eqlab{sin2element}
\eeq
which can be rewritten by
\bseq
\bea
\sin^2 2\theta_{13} &=& 4 |U_{e3}|^2 \l(1-|U_{e3}|^2\r)\,, 
\eqlab{sin2_13} \\
\sin^2 2\theta_{23} &=& 4 \dfrac{U_{\mu 3}^2}{1-|U_{e3}|^2}
                      \l(1-\dfrac{U_{\mu 3}^2}{1-|U_{e3}|^2}\r)\,,
\eqlab{sin2_23} \\
\sin^2 2\theta_{12} &=& 4 \dfrac{U_{e2}^2}{1-|U_{e3}|^2}
                              \l(1-\dfrac{U_{e2}^2}{1-|U_{e3}|^2}\r)\,,
\eqlab{sin2_12}
\eea
\eqlab{sin2_2theta}
\eseq
\hspace*{-2ex}
where $\theta_{ij}$ is the mixing angle between the $i$-th and
$j$-th generations.
In general,
 these mixing angles
are not the same as those obtained from
the two generation analyses of the experimental results.

\subsection{The MNS matrix at the weak scale}
Now let us decide the values of $U_{e2}$, $U_{e3}$ and $U_{\mu3}$,
which are independent parameters of the MNS matrix,
from the data of
neutrino-oscillation experiments{\footnote
{Neutrino-oscillation probabilities
are shown in Appendix~\ref{app:MNS}.}
}.

\subsubsection{$U_{e3}$}
 The CHOOZ experiment \cite{CHOOZ} measures the survival
probability of $\overline{\nu_e^{}}$.
 The result of this experiment shows
\beq
\sin^2 2 \theta_{\rm CHOOZ}
< 0.18\,,
~~~\mbox{for}~~~ \delta m_{\rm CHOOZ}^2 > 1 \times 10^{-3} \mbox{eV}^2.
\eqlab{chooz0}
\eeq
From \eqref{surv}, we can obtain the following constraint
\bseq
\bea
\l | U_{e3} \r|^2 \l( 1-\l| U_{e3} \r|^2 \r)
&<& 0.045\,,
\eqlab{region} \\
\mbox{for  \quad}
|m_3^2-m_1^2| \simeq |m_3^2-m_2^2| 
&>& 1 \times 10^{-3} \mbox{eV}^2\,.
\eqlab{CHOOZ-MASS}
\eea
\eseq

\noindent
In this article we assume the (1,3) element
of the MNS matrix at the weak scale $m_z^{}$ to be
\beq
U_{e3} = 0\,.
\eqlab{assUe3}
\eeq

\subsubsection{$U_{\mu 3}$}
The atmospheric neutrino data
\cite{Atm4,SK4} suggest the maximal mixing
of $\nu_\mu^{}\to\nu_X^{}$ ($\nu_X^{} \neq \nu_\mu^{}, \nu_e^{}$)
oscillation\footnote{{The oscillation of $\nu_\mu^{}\to\nu_e^{}$ is not
only disfavored by the CHOOZ experiment data of \eqref{chooz0},
but also disfavored by the Super-Kamiokande data by itself
\cite{Atm4,SK4}.}}.
In the three-flavor analysis, 
the most reliable interpretation of the data is
$\nu_\mu \to \nu_\tau$ oscillation under the condition of
\eqref{A-massd}.
From \eqsref{sin2_23}, (\ref{eqn:assUe3}) and (\ref{eqn:surv}),
we can obtain
\beq
\sin^2 2\theta_{\rm ATM} = 
4 U_{\mu 3}^2 \l(1-U_{\mu 3}^2\r)\,.
\eqlab{ATM2}
\eeq
Thus, by using \eqref{angle-ATM}
we determine the (2,3) element of the MNS matrix at the $m_z^{}$ scale as
\beq
U_{\mu 3} = \dfrac{1}{\sqrt{2}}\,.
\eqlab{assUm3}
\eeq

\subsubsection{$U_{e2}$}
Deficits of solar neutrinos observed at several telestial experiments
\cite{solar4} have been interpreted as 
$\nu_e^{}\to\nu_X^{}$ ($\nu_X^{}\neq\nu_e^{}\,,\overline{\nu_e^{}}$)
oscillation of the three solutions, which are
the MSW small-mixing solution (MSW-S),
the MSW large-mixing solution (MSW-L) \cite{MSW}, and
the vacuum oscillation solution (VO) \cite{Vacuum}.
By using \eqref{assUe3},
the survival probability of $\nu_e^{}$
in \eqref{A9} is simplified as
\beq
P_{\nu_e \to \nu_e} =
1
-4 |U_{e 1}|^2 |U_{e 2}|^2 
\sin^2
\l(
{\dfrac{\delta m_{12}^2}{4E}L}
\r)\,.
\eqlab{A9-2-e}
\eeq
\Eqref{assUe3} also simplifies \eqref{sin2_12} to be
\bea
\sin^2 2\theta_{\rm SUN}
 &=& 4U_{e2}^2 \l( 1 -U_{e2}^2 \r) \,.
\eea
Thus,
from \eqsref{angle-SUN},
we determine the value of $U_{e2}$ at the $m_z^{}$ scale as
\beq
\mbox{{ MSW-S}~:~}U_{e2}=0.0042\,,
\mbox{\quad}
\mbox{{ MSW-L}~:~}U_{e2}=\dfrac{1}{\sqrt{2}}\,,
\mbox{\quad}
\mbox{{ VO}~:~}U_{e2}=\dfrac{1}{\sqrt{2}}\,,
\eqlab{solar-mass}
\eeq
 for each solution.

\subsubsection{The MNS matrix at $m_z^{}$ scale}

In this article we neglect
Majorana phases of $\varphi_{2,3}$ in the MNS matrix of
\eqref{MNS1} for simplicity.
By using \eqsref{MNS-RC},
\eqvref{assUe3} and \eqvref{assUm3},
the MNS matrix at the $m_z^{}$ scale is
determined as
\beq
U_{\rm MNS}=
\bmaT
      {\cos \theta}           &   \sin \theta                 & 0 \\
& & \\
\dfrac{-\sin \theta}{\sqrt{2}}& \dfrac{\cos \theta}{\sqrt{2}} &
\dfrac{1}{\sqrt{2}}\\
& & \\
\dfrac{\sin \theta}{\sqrt{2}}& \dfrac{-\cos \theta}{\sqrt{2}} &
\dfrac{1}{\sqrt{2}}
\emaT\,,
\eqlab{MNS2}
\eeq
where $\theta$\footnote{{From now on,
$\theta_{12}$ is denoted by $\theta$ 
unless we note explicitly.}} 
depends on the solution of the solar neutrino deficits
as
\beq
\sin \theta =
\l\{
\begin{array}{cll}
0.0042 & (\theta=0.0042) & \mbox{({MSW-S})}\,,\\
\dfrac{1}{\sqrt{2}}& (\theta=\dfrac{\pi}{4}) & \mbox{({MSW-L})}\,,\\
\dfrac{1}{\sqrt{2}}& (\theta=\dfrac{\pi}{4}) & \mbox{({VO})}\,,\\
\end{array}
\r.
\eqlab{ass-12}
\eeq
which are obtained from \eqsref{solar-mass}.

\section{Quantum Corrections of $\kappa$}
\label{sec:RGE}
\setcounter{equation}{0}
 The renormalization group equation (RGE) of $\kappa$,
which is the coefficient of dimension-five operator 
in the effective Lagrangian of \eqref{W_low},
has been studied in Refs.~\cite{nue-RGE1,nue-RGE2}.
 It is expected that $\kappa$ is produced
by the  see-saw mechanism \cite{seesaw}
at the high energy-scale $M_R^{}$.
 In the MSSM, 
$\kappa$ satisfies the following
RGE at the one-loop level \cite{nue-RGE2}:
\beq
8\pi^2\dfrac{d}{dt}\kappa =
      \l\{ \tr \l( 3\hu\hu^{\dagger} \r) 
       - 4\pi \l( 3\alpha_2 + \dfrac{3}{5} \alpha_1 \r) \r\} \kappa
+\dfrac{1}{2} \l\{ \l( \he \he^{\dagger}\r) \kappa
                 + \kappa\l( \he \he^{\dagger}\r)^{T} \r\}\,,
\eqlab{RGE_kappa}
\eeq
where $t=\ln \mu$, and $\mu$ is the renormalization scale.
$\hu$ is the up-quark Yukawa matrix.
 We notice that once $\he$ is taken diagonal,
this form is kept thought all energy-scale
at the one-loop level.
 It is because there are no lepton-flavor-mixing terms
except for $\kappa$ in the MSSM Lagrangian.
 In this base $\kappa$ is diagonalized by the MNS matrix,
and \eqref{RGE_kappa} is simplified to be
\beq
8\pi^2\dfrac{d}{dt} \ln \kappa_{ij} =
      \tr \l( 3\hu\hu^{\dagger} \r) 
       - 4\pi \l( 3\alpha_2 + \dfrac{3}{5} \alpha_1 \r) 
+\dfrac{1}{2} \l( y_i^2 + y_j^2 \r)\,,
\eqlab{RGE_kappa2}
\eeq
where $y_i$ ($i=1 \sim 3$) stands for the $i$-th generation
charged-lepton Yukawa coupling. 

 The RGE of $\kappa$ has two important features\footnote{{
More detail discussions are in the Ref.~\cite{no7}.}}.
 One is that none of the phases in $\kappa$ depend on
the energy-scale.
 The other is that the RGE of $\kappa$ can be governed by only
$n_g^{}$ equations, where $n_g^{}$ stands for the generation number.
 In the three generation case ($n_g^{}=3$),
$\kappa$ can be parameterized as
\bea
\kappa &=& \kappa^{}_{33}
\bmaT
r^{}_1 & c^{}_{12} \sqrt{r_1 r_2} & c^{}_{13} \sqrt{r_1} \\
c^{}_{12} \sqrt{r_1 r_2} & r_2 & c^{}_{23} \sqrt{r_2} \\
c^{}_{13} \sqrt{r_1} & c^{}_{23} \sqrt{r_2} & 1
\emaT\,,
\nn \\
&=& \kappa^{}_{33}
\bmaT
\sqrt{r_1^{}} &          0   &   0 \\
0             &\sqrt{r_2^{}} &   0 \\
0             &  0           &   1 \\
\emaT
\bmaT
1        & c^{}_{12} & c^{}_{13} \\
c^{}_{12}& 1         & c^{}_{23} \\
c^{}_{13}& c^{}_{23} & 1
\emaT
\bmaT
\sqrt{r_1^{}} &          0   &   0 \\
0             &\sqrt{r_2^{}} &   0 \\
0             &  0           &   1 \\
\emaT\,.
\eqlab{def_kappa}
\eea
Here $c_{ij}$s ($i = 1,2$ and $j = 2,3$)
are defined as
\beq
 c_{ij}^2 \equiv \dfrac{\kappa_{ij}^2}{\kappa_{ii}^{} \kappa_{jj}^{}}\,,
\eeq
which are energy-scale independent complex parameters.
$r_i^{}$s in \eqref{def_kappa} are defined as
\beq
r_i \equiv \dfrac{\kappa^{}_{ii}}{\kappa^{}_{33}}\,, 
\mbox{\qquad} (i=1,2)\,,
\eqlab{def_ki}
\eeq
which are always taken real and non-negative
by the redefinitions of lepton fields.
Since the MNS matrix is independent of the overall factor $\kappa_{33}$,
only two real parameters, $r_1^{}$ and $r_2^{}$, 
determine the energy-scale dependence of the MNS matrix.

 From \eqref{RGE_kappa2}, the RGE of $r_i$ is given by 
\beq
\dfrac{d}{dt} \ln r_i^{} 
=\dfrac{d}{dt} \ln \dfrac{\kappa^{}_{ii}}{\kappa^{}_{33}} 
= -\dfrac{1}{8\pi^2}\l(y_{\tau}^2-y_{i}^2 \r)\,,
\mbox{\quad}
(i=1,2)\,,
\eqlab{RGE_kr}
\eeq
 where $y_\tau$ is Yukawa coupling of $\tau$.
 Since the right-hand side of \eqref{RGE_kr} is always negative,
the value of $r_i$ decreases as the energy-scale increases.
 \Eqref{RGE_kr} can be solved as
\beq
r_i^{}(M_R^{}) = r_i^{} (m_z) \dfrac{I_i^{}}{I_\tau^{}}\,,
\eqlab{RGE_kr2}
\eeq
where $I_i$ ($i=1(e),2(\mu),\tau$) is defined as
\beq
I_i \equiv 
\exp\l({\dfrac{1}{8\pi^2}{\mbox{ }}
{\displaystyle{\int}
 _{\scriptstyle \ln \l(m_z^{}\r)}
 ^{\scriptstyle \ln \l(M_R^{}\r)} y^2_i {\mbox{ }} dt }}\r)\,.
\eqlab{Ii}
\eeq
From \eqsref{def_kappa} and \eqvref{RGE_kr2}, 
$\kappa$ at the $M_R^{}$ scale is determined as 
\bea
\kappa(M_R) 
&=&
\dfrac{\kappa_{33}(M_R)}{\kappa_{33}(m_z)}
\bmaT
\sqrt{I_e/I_\tau} &          0         &   0 \\
0                 &\sqrt{I_\mu/I_\tau} &   0 \\
0                 &  0                 &   1 \\
\emaT
{\kappa(m_z)}
\bmaT
\sqrt{I_e/I_\tau} &          0         &   0 \\
0                 &\sqrt{I_\mu/I_\tau} &   0 \\
0                 &  0                 &   1 \\
\emaT\,.
\eqlab{def_kappa2}
\eea
 We discuss cases of $\kappa_{33}=0$ and
$\kappa_{11}=\kappa_{22}=\kappa_{33}=0$ in Appendix~\ref{app:RGE-K}.

 Now let us define small parameters of
\beq
\epsilon_{e,\mu} = 1 -\sqrt{\dfrac{I_{e,\mu}}{I_\tau}}
\eqlab{def_eps}
\eeq
for the later discussions.
\Fgsref{epsilon_e} and \Fgvref{epsilon_mu} show  
$\tan \beta (= {\vev{\phi_{\Su}}}/{\vev{\phi_{\Sd}}})$
dependences of $\epsilon_{e,\mu}$
with $M_R^{}=10^{13}$GeV and $10^{6}$GeV.
They show that magnitudes of $\epsilon_{e,\mu}$
increase as $\tan \beta$ increases.
This is because the value of $\sqrt{I_i/I_\tau}$ can be mainly
determined by $I_\tau$, and the quantum correction from $\tau$
becomes large in the large $\tan \beta$ region{\footnote{
We show the approximation of $\sqrt{I_i/I_\tau}$
in Appendix~\ref{app:Itau}.
}}.
\begin{figure}[htb]
\begin{center}
 {\scalebox{1.}{\includegraphics{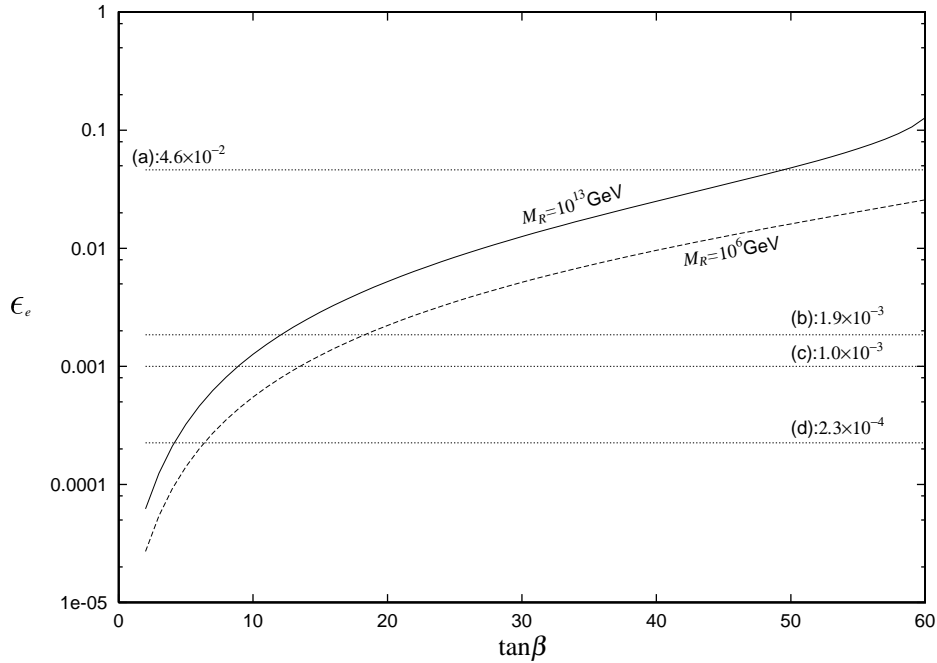}}  } 
\end{center}
 \vspace{-1.5em}
 \caption{$\tan \beta$ dependence of $\epsilon_e$.
Solid-line (dashed-line) shows
$M_R^{}=10^{13}$GeV ($10^6$GeV).
Each dotted-line shows 
 (a):$\numt{4.6}{-2}$, (b):$\numt{1.9}{-3}$,
 (c):$\numt{1.0}{-3}$ and (d):$\numt{2.3}{-4}$.
}
 \fglab{epsilon_e}
\end{figure}
\begin{figure}[hbt]
\begin{center}
 {\scalebox{1.}{\includegraphics{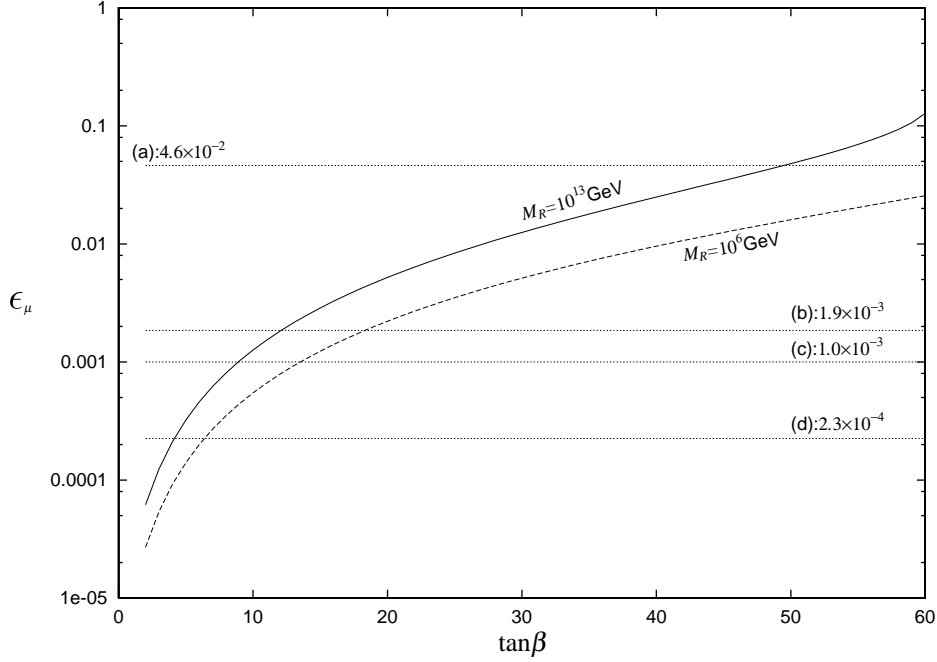}} }
\end{center}
 \vspace{-1.5em}
 \caption{$\tan \beta$ dependence of $\epsilon_\mu$.
Solid-line (dashed-line) shows
$M_R^{}=10^{13}$GeV ($10^6$GeV).
Each dotted-line shows 
 (a):$\numt{4.6}{-2}$, (b):$\numt{1.9}{-3}$,
 (c):$\numt{1.0}{-3}$ and (d):$\numt{2.3}{-4}$.}
 \fglab{epsilon_mu}
\end{figure}
We notice that the 
large $\tan \beta$ means large quantum corrections
of $\epsilon_{e,\mu}$.
 Solid-lines (dashed-lines)
in \Fgsref{epsilon_e} and \Fgvref{epsilon_mu}
stand for the $\tan \beta$ dependences of $\epsilon_{e,\mu}$ 
with $M_R^{}=10^{13}$GeV ($10^{6}$GeV).
 The value of $\epsilon_{e,\mu}$ with $M_R^{}=10^6$GeV
is smaller than that with $M_R^{}=10^{13}$GeV
at the same value of $\tan \beta$.
 This is because charged-lepton Yukawa couplings
are enhanced at the high energy scale.
 Hereafter, we fix the $M_R^{}$ scale at $10^{13}$GeV
in our numerical analyses.

\section{Two Generation Neutrinos}
\label{sec:Toy}
\setcounter{equation}{0}
 In this section, 
we neglect the first generation contributions for simplicity,
and discuss the stability
of the MNS matrix against quantum corrections 
in the two generation neutrinos.

Neglecting the first generation,
\eqref{def_kappa2} becomes
\bea
\kappa(M_R) 
&=&
\dfrac{\kappa_{33}(M_R)}{\kappa_{33}(m_z)}
\bma{cc}
\sqrt{I_\mu/I_\tau} &   0 \\
0                   &   1 \\
\ema
{\kappa(m_z)}
\bma{cc}
\sqrt{I_\mu/I_\tau} &   0 \\
  0                 &   1 \\
\ema\,.
\eqlab{def_kappa2gen}
\eea
We parameterize $\kappa(m_z^{})$ as
\beq
\kappa (m_z) =
\bma{cc}
 \cos \theta_{23} & \sin \theta_{23} \\
-\sin \theta_{23} & \cos \theta_{23}
\ema
\bma{cc}
 {\kappa_2} & 0 \\
  0  & {\kappa_3} 
\ema
\bma{cc}
 \cos \theta_{23} & -\sin \theta_{23} \\
 \sin \theta_{23} &  \cos \theta_{23}
\ema\,,
\eqlab{2gen-kappa}
\eeq
where
$\kappa_{2,3}$ are eigenvalues of $\kappa$ at the
$m_z^{}$ scale.
From \eqref{def_kappa2gen},
the mixing angle at the $M_R$ scale
$(\hat{\theta}_{23})$ is given by
\bea
\tan 2 \hat{\theta}_{23}
&=&
\dfrac{\delta k \sin 2 \theta_{23}  \l( 1 - \epsilon \r)}
{\delta k \cos 2 \theta_{23}  
 + \epsilon \l(2-\epsilon \r)\l( {\kappa_2} + \delta k \sin^2 \theta_{23} \r) }
\,,
\eqlab{def_23R}
\eea
where
\beq
\epsilon \equiv \epsilon_{\mu}\,,
\mbox{\quad and \quad}
\delta k \equiv {\kappa_3} - {\kappa_2}\,.
\eeq
Hereafter we denote the mixing angles at the $M_R^{}$ scale
as $\hat{\theta}_{ij}$s.
\Fgref{epsilon_mu} shows $0<\epsilon<0.15$
for $2\leq \tan \beta \leq 60$ with $M_R^{}=10^{13}$ GeV.
 We notice that $|\delta k|$ is not 
necessarily smaller than $|\kappa_{2,3}|$,
because ${\kappa_2}$ can take the opposite sign of $\kappa_3$.
 We classify the neutrino mass hierarchies
into the following three cases as
Type  A$^{(2)}$: $|\kappa_3|\gg|\kappa_2|$,
Type B1$^{(2)}$: $\kappa_{3}\simeq\kappa_2$ and
Type B2$^{(2)}$: $\kappa_3 \simeq -\kappa_2$.
 We study the stability of the mixing angle in each case.

\begin{enumerate}
 \item 
\UL{Type A$^{(2)}$}:

 When $|{\kappa_2}|$ is much smaller than $|{\kappa_3}|$
(\ie $\delta k \sim |\kappa_3| \gg |\kappa_2| \simeq 0$),
\eqref{2gen-kappa} becomes
\bea
\kappa (m_z)
&\simeq&\kappa_3 \c23{^2}
\bma{cc}
 \tan^2 \theta_{23} & \tan \theta_{23} \\
 \tan \theta_{23} & 1
\ema\,.
\eea
In this case,
the mixing angle $\hat{\theta}_{23}$ is
given by
\beq
{\tan 2 {\hat{\theta}_{23}}}
=
\dfrac{\s23{2}(1-\epsilon)}
{\c23{2}+ \epsilon (2-\epsilon) \s23{^2} }
=
{\tan 2 {{\theta}_{23}}}
\l(1-\epsilon \sec 2 \theta_{23}\r) + O(\epsilon^2)
\eqlab{23R-2-typeA}
\eeq
from \eqref{def_23R}.
This means
$\hat{\theta}_{23}$ is stable against the quantum correction of
$\epsilon$.

\item
\UL{Type B1$^{(2)}$}:

In the case of ${\kappa_2 \simeq \kappa_3}$
(\ie $\kappa_2\kappa_3 >0,~0<|\delta k| \ll |\kappa_{2,3}|$),
$\kappa (m_z^{})$ is given by
\beq
\kappa (m_z^{}) =
\bma{cc}
\kappa_3^{} - \dfrac{\delta k}{2} & \dfrac{\delta k}{2} \\
& \\
\dfrac{\delta k}{2}            &\kappa_3^{} - \dfrac{\delta k}{2} 
\ema
\simeq
\l(\kappa_3^{} - \dfrac{\delta k}{2}\r)
\bma{cc}
~~~1~~~ & \dfrac{\delta k}{2 \kappa_3} \\
& \\
\dfrac{\delta k}{2 \kappa_3} & ~~~1~~~\\
\ema\,,
\eqlab{2gen-mass-B1}
\eeq
around the maximal mixing
$\theta_{23}=\pi/4$.
Let us
discuss the stability of \eqref{2gen-mass-B1}
against the renormalization effects.
\Eqref{def_23R} induces
the mixing angle $\hat{\theta}_{23}$ as
\beq
{\tan 2 {\hat{\theta}_{23}}}
\simeq
\dfrac{1}{2\epsilon}~\dfrac{\delta k}{\kappa_3}\,.
\eeq
This means the mixing angle of 
$\hat{\theta}_{23}$ strongly depends on the
quantum correction of $\epsilon$.
 When the value of $\delta k/\kappa_3$ is larger than $\epsilon$,
the mixing angle
does not receive significant change from 
renormalization corrections.
 On the other hand,
when $\epsilon > \delta k/\kappa_3^{}$,
the mixing angle strongly depends on RGE effects, and
the mixing angle at the $M_R^{}$ scale can be small
even if the maximal mixing is realized at the $m_z^{}$ scale.
This situation has been already discussed in Ref.~\cite{no5}.

\item
\UL{Type B2$^{(2)}$}:

If the absolute value of $\kappa_{2,3}$ are the same order,
but they have opposite signs from each other
(\ie $\kappa_2 \kappa_3<0$, $|\delta k| \simeq 2 |\kappa_{2,3}|$),
$\kappa(m_z^{})$ is given by
\beq
\kappa(m_z^{})
\simeq{\kappa_3 \c23{2}}
\bma{cc}
-1 & \tan 2 \theta_{23} \\
& \\
\tan 2 \theta_{23} & 1
\ema\,.
\eeq
In this case, \eqref{def_23R} induces
\beq
\tan 2 \hat{\theta}_{23}
=
\tan 2 {\theta}_{23}
\l( \dfrac{2\l(1-\epsilon\r)}{2-\epsilon\l(2-\epsilon\r)} \r)
=
\tan 2 {\theta}_{23} + O(\epsilon^2)\,.
\eeq
This means that the mixing angle is stable against
the small change of $\epsilon$. 
\end{enumerate}

\begin{table}[htb]
\begin{footnotesize}
\begin{center}
\begin{tabular}{|c|c|c|c|}
\hline
hierarchy& $\kappa(m_z^{})$ 
& $\tan 2 \hat{\theta}_{23}$(mixing angle at the $M_R^{}$) &Stability\\
\hline
& & & \\
$
\begin{array}{l}
\mbox{Type A}^{(2)} \\
(\kappa_3^{}\gg\kappa_2^{})\\
\end{array}
$
&
$
\kappa_3^{} \c23{^2}
\bma{cc}
 \tan^2 \theta_{23} & \tan \theta_{23}\\
\\
 \tan \theta_{23}& 1
\ema
$ & $ 
{\tan 2 {{\theta}_{23}}}
\l(1-\epsilon \sec 2 \theta_{23}\r) + O(\epsilon^2)
$&stable\\
& & & \\
\hline
& & & \\
$
\begin{array}{l}
\mbox{Type B1}^{(2)} \\
(\kappa_3^{}\simeq\kappa_2^{}) \\
(\theta_{23}=\pi/4)
\end{array}
$
&
$
\l(\kappa_3^{} - \dfrac{\delta k}{2}\r) 
\bma{cc}
1 & \dfrac{\delta k}{2\kappa_3} \\
& \\
\dfrac{\delta k}{2\kappa_3} &1 
\ema
$ & $
\simeq
\dfrac{1}{2\epsilon}~\dfrac{\delta k}{\kappa_3}
$& 
$
\begin{array}{c}
\mbox{unstable}\\
(\epsilon > \dfrac{\delta k}{\kappa_3}) \\
 \end{array}
$\\
& & & \\
\hline
& & & \\
$
\begin{array}{l}
\mbox{Type B2}^{(2)} \\
(\kappa_3^{}\simeq -\kappa_2^{}) \\
\end{array}
$
&
$
{\kappa_3^{}\c23{2}}
\bma{cc}
-1 & \tan 2 \theta_{23} \\
\\
\tan 2 \theta_{23} & 1
\ema
$ & $
\tan 2 {\theta}_{23} + O(\epsilon^2)
$&stable\\
& & & \\
\hline
\end{tabular}
\end{center}
\caption{Stabilities of the two generation case.}
\tblab{2genSum}
\end{footnotesize}
\end{table}
\Tbref{2genSum} shows
the stability of the MNS matrix
in the two generation neutrinos.
Mixing angles of 
Type A$^{(2)}$ and Type B2$^{(2)}$
are stable against the small change of $\epsilon$.
This means they are stable against 
quantum corrections.
The mixing angle of 
Type B1$^{(2)}$ is unstable around $\theta_{23}=\pi/4$
when $\epsilon > \delta k/\kappa_3$.
The mixing angle of this case
is sensitive to the quantum corrections, and
determined from the ratio of $\delta k / \kappa_3$ and $\epsilon$.
\Tbref{2genSum} shows
that the neutrino mass hierarchy and the relative sign assignment
of mass eigenvalues play the crucial roles
for the stability of the MNS matrix.

\section{Three Generation Neutrinos}
\label{sec:3gen2}
\setcounter{equation}{0}

 Now let us discuss the stability of the MNS
matrix in the three generation neutrinos
against quantum corrections
according to the classification of 
mass hierarchies in \eqsref{type}.

\subsection{Type A}
In Type A,
the mass spectrum is given by
\beq
m_1  =  0\,,
\mbox{\quad}
m_2  =  \sqrt{\Delta m_{\rm solar}^2}\,,
\mbox{\quad}
m_3  =  \sqrt{\Delta m_{\rm solar}^2 + \Delta m_{\rm ATM}^2}\,.
\eqlab{mass-A}
\eeq
 Neutrino masses of this type have large hierarchies. 
 In this type, 
there are following relative sign assignments for mass eigenvalues.
\bseq
\bea
\mbox{case (a1): \quad} m_{\nu}^{\rm a1} &=& diag.(0, m_2^{},m_3^{})
\eqlab{ma1} \\
\mbox{case (a2): \quad}m_{\nu}^{\rm a2} &=& diag.(0,-m_2^{},m_3^{})
\eqlab{ma2}
\eea
\eqlab{typea}
\eseq

\vspace{-3ex}

\noindent
The neutrino mass matrices at the weak scale
for (a1) and (a2) are shown in \Tbref{typeA},
where we write the leading order of each element, and
the small parameter $\xi_a^{}$ is defined as
\beq
\xi^{}_{\rm a} = \dfrac{m_2^{}}{m_3^{}}
\simeq \sqrt{\dfrac{\Delta m^2_{\rm solar}}{\Delta m^2_{\rm ATM}}}\,.
\eqlab{def_xi_a}
\eeq

The relation between the neutrino mass matrices
at $m_z^{}$ and $M_R^{}$ is given by
\beq 
M_\nu^{}(M_R^{})=
\bmaT
1-\epsilon_e &     0         &   0 \\
0            &1-\epsilon_\mu &   0 \\
0            &  0            &   1 \\
\emaT
M_{\nu}^{}(m_z^{}) 
\bmaT
1-\epsilon_e &          0    &   0 \\
0            &1-\epsilon_\mu &   0 \\
0            &  0            &   1 \\
\emaT\,.
\eqlab{parts-mr}
\eeq
The mixing angles of the MNS matrix at $M_R^{}$ 
can be obtained from \eqref{parts-mr}.
\begin{table}[htb]
\bsm
\begin{center}
\begin{tabular}{|c|c|c|}
\cline{2-3}
\multicolumn{1}{c|}{} &{neutrino mass matrix up to the leading order} & Stability \\
\hline
 & &  \\
$
\begin{array}{c}
\mbox{case(a1)} \\
diag.(0, m_2^{},m_3^{})
\end{array}
$
&
$
\dfrac{m_3^{}}{2}
\bmaT
2 \xi^{}_{\rm a} \sin^2 \theta 
 &  \xi^{}_{\rm a} \sin 2 \theta/\sqrt{2} 
 & -\xi^{}_{\rm a} \sin 2 \theta/\sqrt{2}\\
 \xi^{}_{\rm a} \sin 2 \theta /\sqrt{2} & 1 & 1 \\
-\xi^{}_{\rm a} \sin 2 \theta /\sqrt{2} & 1 & 1
\emaT$
& stable \\
 & &  \\
\hline
 & &  \\
$
\begin{array}{c}
\mbox{case(a2)} \\
diag.(0,-m_2^{},m_3^{})
\end{array}
$
&
$
\dfrac{m_3^{}}{2}
\bmaT
 -2 \xi^{}_{\rm a} \sin^2 \theta 
 &-\xi^{}_{\rm a} \sin 2 \theta /\sqrt{2} 
 & \xi^{}_{\rm a} \sin 2 \theta /\sqrt{2}\\
-\xi^{}_{\rm a} \sin 2 \theta /\sqrt{2} & 1 & 1 \\
 \xi^{}_{\rm a} \sin 2 \theta /\sqrt{2} & 1 & 1
\emaT$
& stable \\
 & &  \\
\hline
\end{tabular}
 \caption{Neutrino mass matrices at the weak scale for (a1) and (a2).}
 \tblab{typeA}
\end{center}
\esm
\end{table}
\begin{table}[htbp]
\bsm
\begin{center}
\begin{tabular}{|c|c|c|c|c|}
\cline{3-5}
\multicolumn{1}{c}{} & \multicolumn{1}{c|}{} & MSW-S & MSW-L & VO \\
\hline
     & $\sin^2 2\hat{\theta}_{12}$
        & $0.007 \sim 0.006 $ & $1 \sim 0.996$ & $1 \sim 0.996$ \\
\cline{2-5}
(a1) & $\sin^2 2\hat{\theta}_{23}$
        & $1 \sim 0.98 $  & $1 \sim 0.98$ & $1 \sim 0.98$ \\
\cline{2-5}
     & $\sin^2 2\hat{\theta}_{13}$
        &$0.0 \sim \numt{2.7}{-7}$  & $0.0 \sim \numt{8.8}{-5}$ 
                                    & $0.0 \sim \numt{3.7}{-10}$ \\
\hline
\hline
     & $\sin^2 2\hat{\theta}_{12}$
        & $0.007 \sim 0.006 $ & $1 \sim 0.996$ & $1 \sim 0.996$ \\
\cline{2-5}
(a2) & $\sin^2 2\hat{\theta}_{23}$
        & $1 \sim 0.98 $ & $1 \sim 0.98$  & $1 \sim 0.98$ \\
\cline{2-5}
     & $\sin^2 2\hat{\theta}_{13}$
         &$0.0 \sim \numt{2.2}{-7}$  & $0.0 \sim \numt{6.8}{-5}$ 
                                     & $0.0 \sim \numt{3.7}{-10}$ \\
\hline
\end{tabular}
 \caption{
$\tan \beta$ dependences of the mixing angles at $M_R^{}$
in (a1) and (a2) ($2 \leq \tan \beta \leq 60$).}
 \tblab{typeA-mixing}
\end{center}
\esm
\end{table}
 \Tbref{typeA-mixing} shows 
$\tan \beta$ dependences of the mixing angles at $M_R^{}$
in the region of $2\leq \tan \beta \leq 60$.
 Since $U_{e3}\simeq0$ from the numerical analysis,
 the MNS matrix can be regarded as two sets of two generation mixings,
which are between the first and the second generations
and also between the second and the third generations.
 Since the large mass hierarchies exist in
the first and the second generations
and also in the second and the third generations
as $m_1^{}\ll m_2^{}$ and $m_2^{}\ll m_3^{}$,
$\sin^2 2\hat{\theta}_{12}$ and $\sin^2 2\hat{\theta}_{23}$
are also stable against quantum corrections
on the analogy of Type A$^{(2)}$.
The numerical analyses really show 
all mixing angles are not changed significantly
by quantum corrections.
This means that the MNS matrix is stable against 
quantum corrections in (a1) and (a2).
\subsection{Type B}
At first
we consider the case that $m_1^{}$ is larger than $m_2^{}$.
Then the mass spectrum of this type is given by
\beq
m_1  =  \sqrt{\Delta m_{\rm ATM}^2 + \Delta m_{\rm solar}^2}\,, \mbox{\quad}
m_2  =  \sqrt{\Delta m_{\rm ATM}^2}\,, \mbox{\quad}
m_3  =  0 \,.
\eqlab{mass-B}
\eeq
There are following two cases according to the 
relative sign assignments for masses.
\bseq
\bea
\mbox{case (b1):\quad} m_{\nu}^{b1} &=& diag.(m_1, m_2,0) \eqlab{mb1} \\
\mbox{case (b2):\quad} m_{\nu}^{b2} &=& diag.(m_1,-m_2,0) \eqlab{mb2}
\eea
\eqlab{typeb}
\eseq
\begin{table}
\bsm
\begin{center}
\begin{tabular}{|c|c|c|}
\cline{2-3}
\multicolumn{1}{c|}{} &{neutrino mass matrix up to the leading order} & Stability \\
\hline
 & &  \\
$
\begin{array}{c}
\mbox{case(b1)} \\
diag.(m_1^{},m_2^{},0)
\end{array}
$
&
$
\dfrac{m_1^{}}{2}
\bmaT
2  &  {\xi_b^{}\sin 2\theta / \sqrt{2}}
   & -{\xi_b^{}\sin 2\theta / \sqrt{2}} \\
 {\xi_b^{}\sin 2\theta /\sqrt{2}} &  1 & -1 \\
-{\xi_b^{}\sin 2\theta /\sqrt{2}} & -1 &  1
\emaT$
&
$
\begin{array}{c}
\mbox{unstable} \\
(\sin^2 2 \theta_{12})
\end{array}
$
\\
 & &  \\
\hline
 & &  \\
$
\begin{array}{c}
\mbox{case(b2)} \\
diag.(m_1^{},-m_2^{},0)
\end{array}
$
&
$
-\dfrac{m_1^{} \cos 2 \theta}{2}
\bmaT
 -2 &\sqrt{2}\tan 2 \theta & -\sqrt{2}\tan 2 \theta\\
 \sqrt{2}\tan 2 \theta & 1 & -1 \\
-\sqrt{2}\tan 2 \theta & -1 & 1
\emaT $
& stable \\
 & &  \\
\hline
\end{tabular}
 \caption{Neutrino mass matrices at $m_z^{}$ for (b1) and (b2).}
\tblab{typeB}
\end{center}
\esm
\end{table}

The neutrino mass matrices at $m_z^{}$
for (b1) and (b2) are listed in \Tbref{typeB},
where
the small parameter $\xi_b^{}$ is defined as 
\beq
\xi_b^{} = \dfrac{m_2^{}-m_1^{}}{m_1^{}} =
- \dfrac{1}{2}
~~\dfrac{\Delta m^2_{\rm solar}}{\Delta m_{\rm ATM}^2}\,.
\eqlab{def_xi_b}
\eeq

We analyze the stability of the MNS matrix
for (b1) and (b2) against quantum corrections
by using $M_\nu^{}(M_R^{})$
obtained from \eqref{parts-mr}.
The results of our numerical analyses  
are listed in \Tbref{typeB-mixing}.
Let us see the details in both (b1) and (b2).

\begin{table}[htbp]
\bsm
\begin{center}
\begin{tabular}{|c|c|c|c|c|}
\cline{3-5}
\multicolumn{1}{c}{} & \multicolumn{1}{c|}{} & MSW-S & MSW-L & VO \\
\hline
     & $\sin^2 2\hat{\theta}_{12}$
     & \multicolumn{1}{|c}{}
     & \multicolumn{1}{c}{see \Fgref{TypeB}}  &\multicolumn{1}{c|}{} \\ 
\cline{2-5}
(b1) & $\sin^2 2\hat{\theta}_{23}$
        & $1 \sim 0.98 $  & $1 \sim 0.98$ & $1 \sim 0.98$ \\
\cline{2-5}
     & $\sin^2 2\hat{\theta}_{13}$
        &$0.0$  & $0.0$ & $0.0$ \\
\hline
\hline
     & $\sin^2 2\hat{\theta}_{12}$
        & $0.007$ & $1$ & $1$ \\
\cline{2-5}
(b2) & $\sin^2 2\hat{\theta}_{23}$
        & $1 \sim 0.98 $ & $1 \sim 0.98$  & $1 \sim 0.98$ \\
\cline{2-5}
     & $\sin^2 2\hat{\theta}_{13}$
         &$0.0$  & $0.0$ & $0.0$ \\
\hline
\end{tabular}
 \caption{
$\tan \beta$ dependences of the mixing angles at $M_R^{}$
in (b1) and (b2) ($2 \leq \tan \beta \leq 60$).}
 \tblab{typeB-mixing}
\end{center}
\esm
\end{table}

\bs

\noindent
\UL{case(b1)}:
Eigenvalues of $M_\nu(M_R^{})$ are given by
\bea
\overline{m}_1^{} = m_1^{}\l(1+\xi_b^{}\r)\,, 
\mbox{\quad}
\overline{m}_2^{} = m_1^{}\l(1-3\epsilon\r)\,,
\mbox{\quad}
\overline{m}_3^{} = 0\,,
\mbox{\quad}
(|\xi_b^{}| < 3|\epsilon|)\,,\nn  \\
\overline{m}_1^{} = m_1^{}\l(1-3\epsilon\r)\,, 
\mbox{\quad}
\overline{m}_2^{} = m_1^{}\l(1+\xi_b^{}\r)\,,
\mbox{\quad}
\overline{m}_3^{} = 0\,,
\mbox{\quad}
(|\xi_b^{}| > 3|\epsilon|)\,,
\eea
up to order of $\epsilon$ and $\xi_b$
for any value of $\theta$.

Numerical analyses of \Tbref{typeB-mixing}
suggest $\sin^2 2\hat{\theta}_{13}$ and $\sin^2 2\hat{\theta}_{23}$
are stable against quantum corrections.
This is because there are large hierarchies 
of $m_1^{}\ll m_3^{}$ and $m_2^{} \ll m_3^{}$
on the analogy of Type A$^{(2)}$.
How about $\sin^2 2\hat{\theta}_{12}$ ?
\Fgref{TypeB} shows that
$\sin^2 2\hat{\theta}_{12}$ can be changed by quantum corrections
according to the solar neutrino solutions.
\begin{figure}[htb]
\begin{center}
 \scalebox{1.0}{\includegraphics{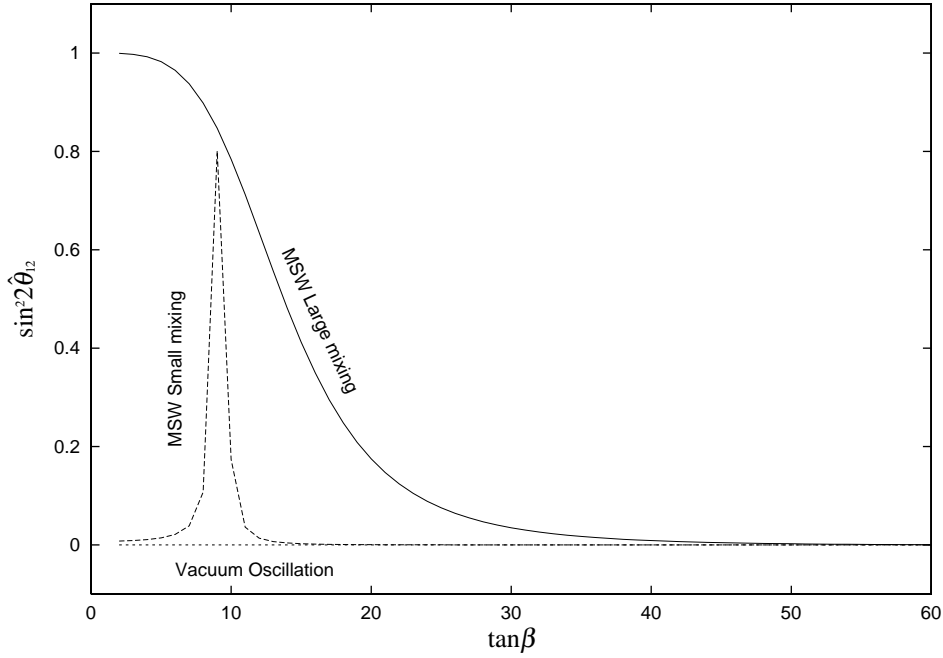}}
 \caption{
$\tan \beta$ dependences of $\sin^2 2 \hat{\theta}_{12}$
according to the solar neutrino solutions 
in (b1).}
 \fglab{TypeB}
\end{center}
\end{figure}

In the MSW-L solution, $\sin^2 2\hat{\theta}_{12}$ is
damping around $\tan \beta \sim 20$.
On the other hand ,
$\sin^2 2 \hat{\theta}_{12} \simeq 0$ even in the small
$\tan \beta$ region in the VO solution.
These situations can be easily understood as follows.
From \eqref{parts-mr},
$\tan 2 \hat{\theta}_{12}$ is estimated to be
\beq
\tan 2 \hat{\theta}_{12} 
\simeq \tan 2 \theta_{12}~
 \l(1-\dfrac{1}{\cos 2 \theta_{12}}
      \dfrac{\epsilon }{|\xi_b^{}|}
       \r)^{-1}\,,
\eqlab{generalB}
\eeq
where we use the approximation of
$\epsilon_e \simeq \epsilon_\mu (\equiv \epsilon)$.
When $\theta_{12}$ is $\pi/4$, \eqref{generalB} 
becomes
\beq
 \tan 2 \hat{\theta}_{12}
\simeq
 \dfrac{\xi_b^{}}{2\epsilon}
\sim
\l\{
\begin{array}{ll}
\dfrac{10^{-3}}{\epsilon}&\mbox{(MSW-L)}\,, \\
& \\
\dfrac{10^{-9}}{\epsilon}&\mbox{(VO)}\,,
 \end{array}
\r.
\eqlab{theta12_B}
\eeq
from \eqref{def_xi_b}.
Dotted-lines of (c) in
\Fgsref{epsilon_e} and \Fgvref{epsilon_mu} show that
$\epsilon$ is much larger than $10^{-3}$ in
the region of $\tan \beta > 20$. 
Then, from \eqref{theta12_B}, 
$\sin^2 2 \hat{\theta}_{12}$ is sufficiently small
when $\tan \beta$ is larger than 20 in the MSW-L solution.
On the other hand, 
\eqref{theta12_B} suggests $\sin^2 2\hat{\theta}_{12} \simeq 0 $
even in the small $\tan \beta$ region 
in the VO solution.

In the MSW-S solution, $\sin^2 2\hat{\theta}_{12}$ has a peak 
around $\tan \beta \sim10$. 
When $\theta_{12} \ll 1$,
\eqref{generalB} becomes
\beq
\tan 2 \hat{\theta}_{12} 
\simeq 2\theta_{12}~
 \l(1-\dfrac{\epsilon }{|\xi_b^{}|}
       \r)^{-1}\,.
\eqlab{theta12_B_MSWS}
\eeq
This means
$\tan 2 \hat{\theta}_{12}$ diverges infinity
($\sin^2 2\hat{\theta}_{12}\sim 1$) at
$\epsilon \simeq |\xi_b^{}|$.
\Eqref{def_xi_b} suggests $|\xi_b^{}|$ is 
about $10^{-3}$.
\Fgsref{epsilon_e} and \Fgvref{epsilon_mu} show
$\epsilon \simeq 10^{-3}$
around $\tan \beta \sim 10$.
 Thus, $\sin^2 2\hat{\theta}_{12}$ becomes one
around $\tan \beta \sim 10$ independently of the value of $\theta_{12}$.
 This is the reason why the peak in the \Fgref{TypeB}
appears in the MSW-S solution.

 Finally, let us see how the stability of the MNS
matrix is changed if we take the mass spectrum as
\beq
m_1  =  \sqrt{\Delta m_{\rm ATM}^2}\,, \mbox{\quad}
m_2  =  \sqrt{\Delta m_{\rm ATM}^2 + \Delta m_{\rm solar}^2}\,, \mbox{\quad}
m_3  =  \sqrt{\Delta m_{\rm solar}^2} \,,
\eqlab{mass-Bp}
\eeq
in stead of \eqsref{mass-B}.
In this case, \eqref{generalB} becomes
\beq
\tan 2 \hat{\theta}_{12} 
\simeq 2 \theta_{12}~
 \l(1+\dfrac{1}{\cos 2 \theta_{12}}
      \dfrac{\epsilon }{|\xi_b^{}|}
       \r)^{-1}\,,
\eqlab{replaced}
\eeq
and \eqref{def_xi_b} is replaced by
\beq
\xi_b^{} = \dfrac{m_2^{}-m_1^{}}{m_1^{}} =
\dfrac{1}{2}
\dfrac{\Delta m^2_{solar}}{\Delta m_{\rm ATM}^2}\,.
\eeq
The behaviors of the mixing angles against quantum corrections
for the MSW-L and the VO solutions are the
same as those of \eqsref{mass-B}.
 As for the MSW-S solution the mixing angle $\hat{\theta}_{12}$
becomes stable against renormalization
effects, and the peak of $\sin^2 2\hat{\theta}_{12}$
around $\tan \beta \sim 10$ disappears
due to the positive sign of \eqref{replaced}.

\bs

\noindent

\UL{case(b2)}:
Eigenvalues of $M_\nu(M_R^{})$ are given by
\bea
\overline{m}_1^{} &=&  m_1^{}\l(\sqrt{1+\xi_b^{}-3\epsilon}
                +\dfrac{1}{2}\l(\xi_b+\epsilon \cos 2 \theta \r)\r)\,, 
\nn \\
\overline{m}_2^{} &=& - m_1^{}\l(\sqrt{1+\xi_b^{}-3\epsilon}
                -\dfrac{1}{2}\l(\xi_b+\epsilon \cos 2 \theta \r)\r)\,, 
\nn \\
\overline{m}_3^{} &=& 0\,,
\eea
up to order of $\epsilon$ and $\xi_b$.
Numerical analyses of \Tbref{typeB-mixing}
show the MNS matrix is stable against quantum corrections.
This can be easily understood as follows.
$\sin^2 2\hat{\theta}_{13}$ and $\sin^2 2\hat{\theta}_{23}$
are stable against quantum corrections,
because there are large hierarchies
between the first and the third generations
and also between the second and the third generations
on the analogy of Type A$^{(2)}$.
 $\sin^2 2\hat{\theta}_{12}$ is also stable against the
renormalization effects,
since the signs of $m_1^{}$ and $m_2^{}$ are different 
from each other
on the analogy of Type B2$^{(2)}$.
Thus, we can conclude
all $\sin^2 2\theta_{ij}$s are stable against 
renormalization effects in case (b2).
\subsection{Type C}
The mass spectrum is given by
\beq
m_1^{} = m_0^{} \,, \mbox{\quad}
m_2^{} = \sqrt{m_0^2+\Delta m_{\rm solar}^2}\,, \mbox{\quad}
m_3^{} = \sqrt{m_0^2+\Delta m_{\rm solar}^2+\Delta m_{\rm ATM}^2}\,,
\eqlab{mass-C}
\eeq
where $m_0^{}$ is the degenerate mass scale.
We take $m_0=1.0$ eV or $0.2$ eV in this article.
 There are following four different cases according to the 
relative sign assignments of neutrino mass eigenvalues.
\bseq
\bea
\mbox{case (c1):\quad} m_{\nu}^{c1} &=&
 diag.(-m_1^{}, m_2^{},m_3^{}) \eqlab{mc1} \\
\mbox{case (c2):\quad} m_{\nu}^{c2} &=&
 diag.( m_1^{},-m_2^{},m_3^{}) \eqlab{mc2} \\
\mbox{case (c3):\quad} m_{\nu}^{c3} &=&
 diag.(-m_1^{},-m_2^{},m_3^{}) \eqlab{mc3} \\
\mbox{case (c4):\quad} m_{\nu}^{c4} &=&
 diag.( m_1^{}, m_2^{},m_3^{}) \eqlab{mc4}
\eea
\eqlab{typec}
\eseq

\vspace*{-3ex}

\noindent
We define the small parameters
\beq
\delta_c^{} \equiv \dfrac{m_3^{}-m_2^{}}{m_0^{}}
         \simeq \dfrac{1}{2}\dfrac{\Delta m^2_{\rm ATM}}{m_0^2}\,,
\mbox{\quad and \quad}
\xi_c^{} \equiv \dfrac{m_2^{}-m_1^{}}{m_0^{}}
         \simeq \dfrac{1}{2}\dfrac{\Delta m^2_{\rm solar}}{m_0^2}\,,
\eqlab{def_xidelta}
\eeq

\noindent
where $\xi_c^{}$ is always smaller than $\delta_c^{}$.
The neutrino mass matrices for all types
are listed in \Tbref{massC} 
up to the leading order for each element.
\begin{table}[htb]
\bsm
\begin{center}
\begin{tabular}{|c|c|c|}
\cline{2-3}
\multicolumn{1}{c|}{} &{neutrino mass matrix up to the leading order} & Stability \\
\hline
 & &  \\
$
\begin{array}{c}
\mbox{case(c1)} \\
diag.(-m_1^{},m_2^{},m_3^{})
\end{array}
$
&
$
\dfrac{m_0^{}}{2}
\bmaT
 {-2\cos 2\theta} & 
 {\sqrt{2}\sin 2 \theta}  &
-{\sqrt{2}\sin 2 \theta} \\
{\sqrt{2}\sin 2 \theta}  &
{1+\cos 2 \theta}&
{1-\cos 2 \theta} \\
-{\sqrt{2}\sin 2 \theta}&
{1-\cos 2 \theta} &
{1+\cos 2 \theta} \\
\emaT
$
&
$
\begin{array}{c}
\mbox{rearrangement}\\
\mbox{between}\\
V_2^{}\mbox{ and }  V_3^{}
\end{array}
$
\\
 & &  \\
\hline
 & &  \\
$
\begin{array}{c}
\mbox{case(c2)} \\
diag.(m_1^{},-m_2^{},m_3^{})
\end{array}
$
&
$
\dfrac{m_0^{}}{2}
\bmaT
 2\cos 2\theta & 
-{\sqrt{2}\sin 2 \theta}  &
 {\sqrt{2}\sin 2 \theta} \\
-{\sqrt{2}\sin 2 \theta}  &
 {1-\cos 2 \theta}&
 {1+\cos 2 \theta} \\
 {\sqrt{2}\sin 2 \theta}&
 {1+\cos 2 \theta} &
 {1-\cos 2 \theta} \\
\emaT
$
&
$
\begin{array}{c}
\mbox{rearrangement}\\
\mbox{between}\\
V_1^{}\mbox{ and }  V_3^{}
\end{array}
$
\\
 & &  \\
\hline
 & &  \\
$
\begin{array}{c}
\mbox{case(c3)} \\
diag.(-m_1^{},-m_2^{},m_3^{})
\end{array}
$
&
$
{m_0^{}}
\bmaT
  -1 &
 -\sqrt{2}\xi_c^{}\sin 2\theta & 
  \sqrt{2}\xi_c^{}\sin 2\theta \\
 -\sqrt{2}\xi_c^{}\sin 2\theta & 
 \delta_c^{}/2 &
 1 \\
 \sqrt{2}\xi_c^{}\sin 2\theta &
 1 &
 \delta_c^{}/2 \\
\emaT
$
&
$
\begin{array}{c}
\mbox{rearrangement}\\
\mbox{between}\\
V_1^{}\mbox{ and }  V_2^{}
\end{array}
$
\\
 & &  \\
\hline
 & &  \\
$
\begin{array}{c}
\mbox{case(c4)} \\
diag.(m_1^{},m_2^{},m_3^{})
\end{array}
$
&
$
{m_0^{}}
\bmaT
 1 &
 \sqrt{2}\xi_c^{}\sin 2\theta & 
-\sqrt{2}\xi_c^{}\sin 2\theta \\
 \sqrt{2}\xi_c^{}\sin 2\theta & 
 1 &
 \delta_c^{}/2 \\
-\sqrt{2}\xi_c^{}\sin 2\theta &
 \delta_c^{}/2 &
 1 \\
\emaT
$
&
$
\begin{array}{c}
\mbox{unstable}\\
\mbox{go to the}\\
\mbox{unit matrix}
\end{array}
$
\\
 & &  \\
\hline
\end{tabular}
 \caption{Neutrino mass matrices at $m_z^{}$ for (c1) $\sim$ (c4).
}
\tblab{massC}  
\end{center}
\esm
\end{table}

In the case of $m_0=1.0$ eV
all solar neutrino solutions of (c3) and (c4),
and the MSW-S solution of (c1) and (c2) 
are excluded by the neutrino-less double-$\beta$
decay experiments \cite{GG},
whose upper limit is given by
 $\vev{m_{\nu_e}} < 0.2$ eV \cite{doublebeta},
where 
\beq
\vev{m_{\nu_e}} = \l|\sum_{i=1}^{3} m_i \l(V_{\rm MNS}\r)_{ei}^2 \r|
=
 \l\{
\begin{array}{ll}
 m_0^{}\l(1+\xi_c^{}\sin^2\theta\r)\,,& \mbox{~~~($m_1^{} m_2^{} >0$)}\,, \\
& \\
 m_0^{}\l(\cos 2\theta -\xi_c^{}\sin^2\theta\r)\,,& 
 \mbox{~~~($m_1^{} m_2^{}<0$)}\,,
\end{array}
\r.
\eqlab{0ne2b-2}
\eeq
from \eqsref{MNS2} and \eqvref{def_xidelta}.
 Thus we analyze the stability of the MNS matrix 
for (c1) and (c2) with $m_0^{}=1.0$ eV,
and for all cases with $m_0^{}=0.2$ eV.
\subsubsection{$m_0=1.0$ eV}
\begin{figure}[htbp]
 \begin{center}
 \scalebox{.62}{\includegraphics{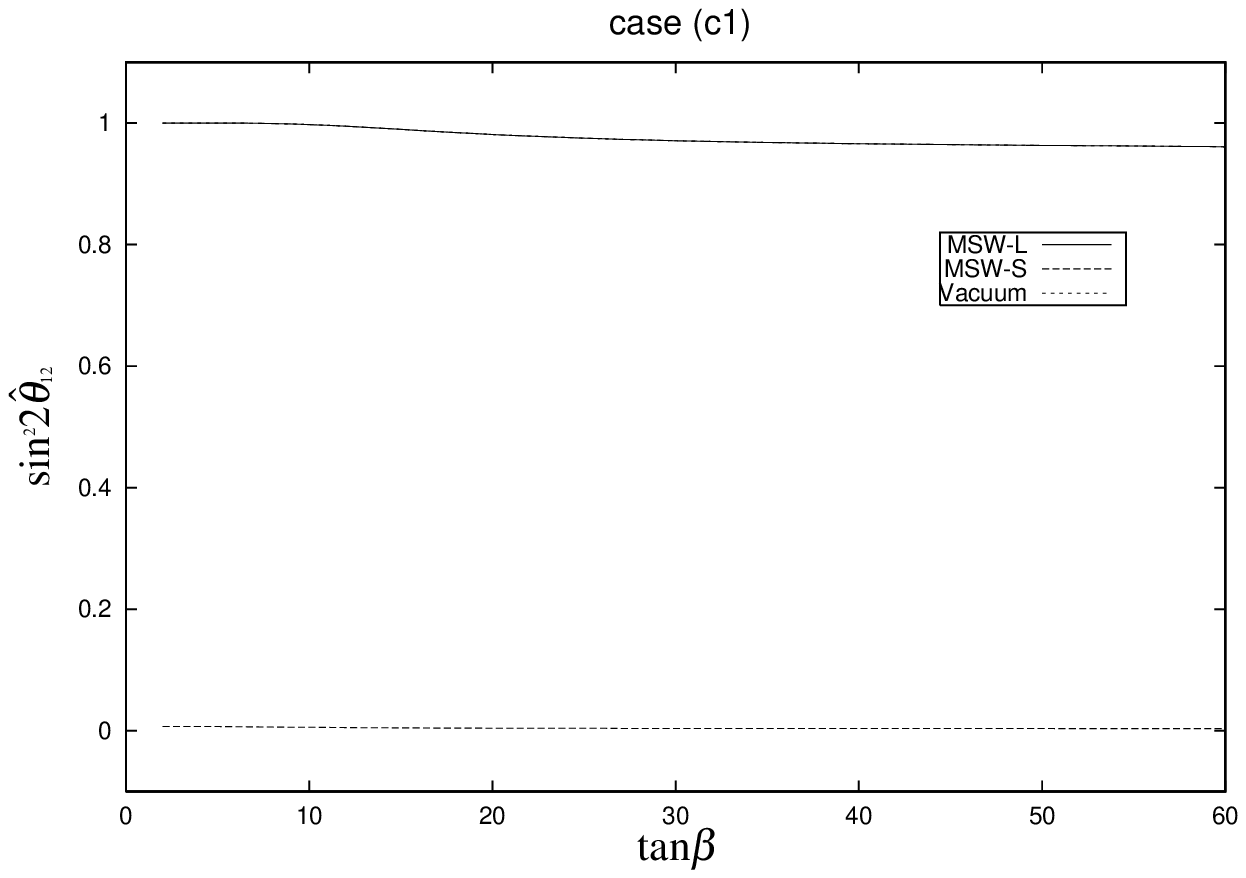}}
 \scalebox{.62}{\includegraphics{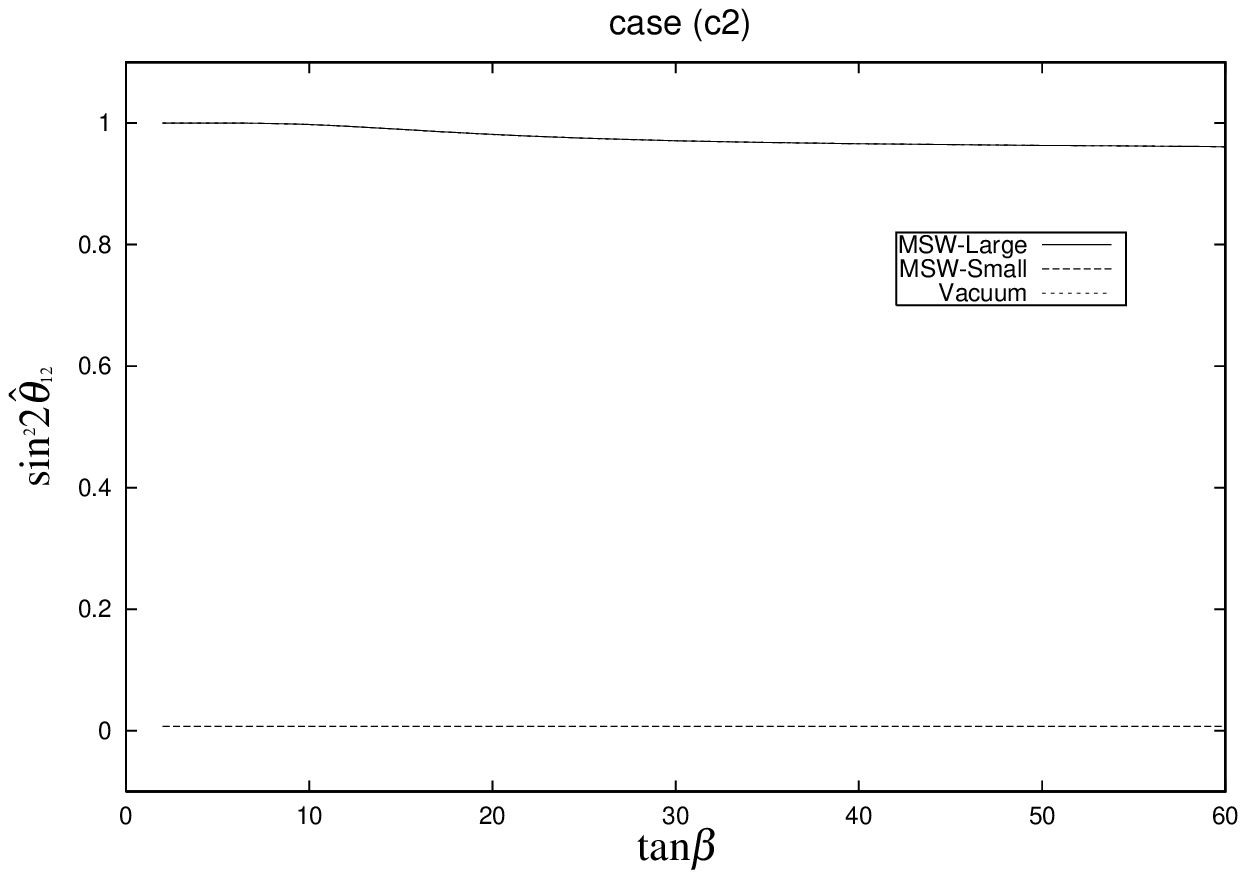}}
 \scalebox{.62}{\includegraphics{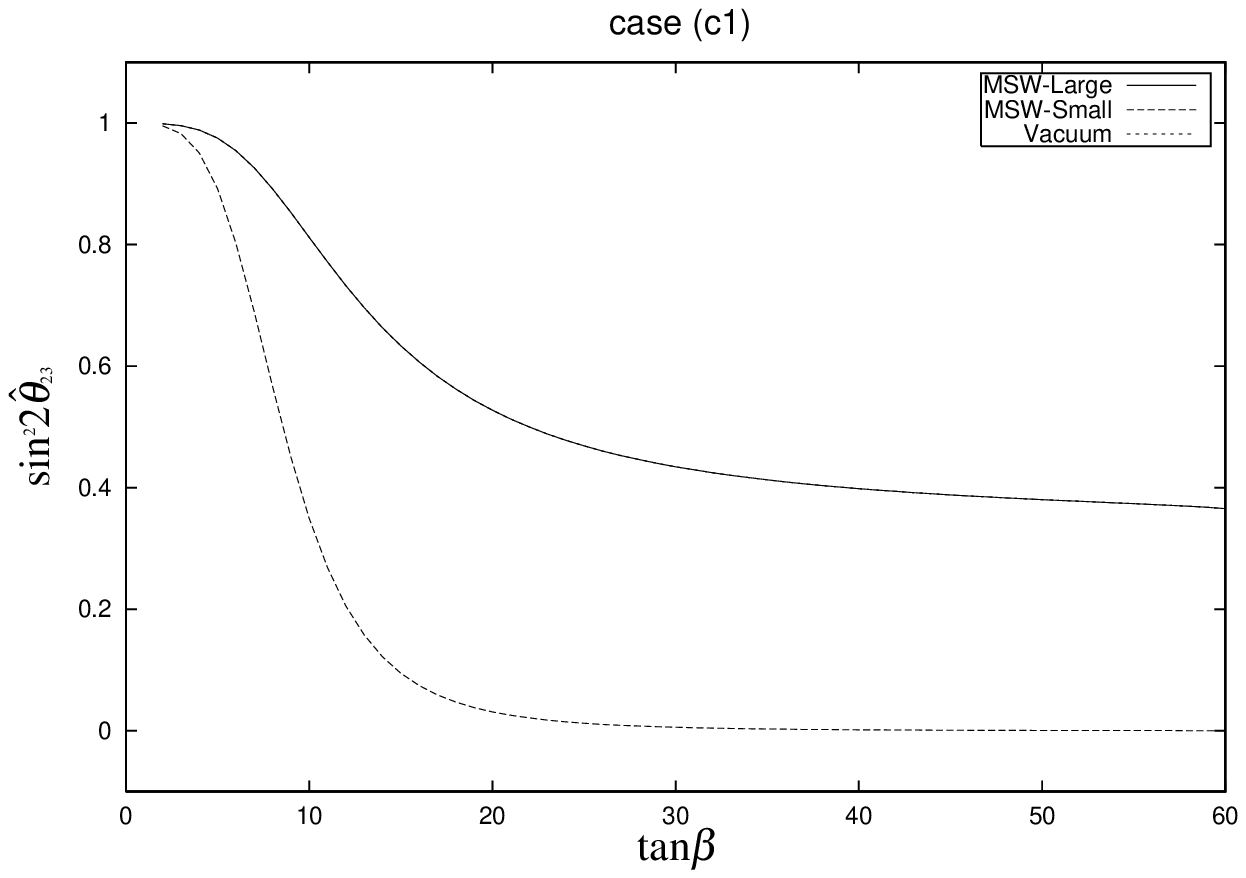}}
 \scalebox{.62}{\includegraphics{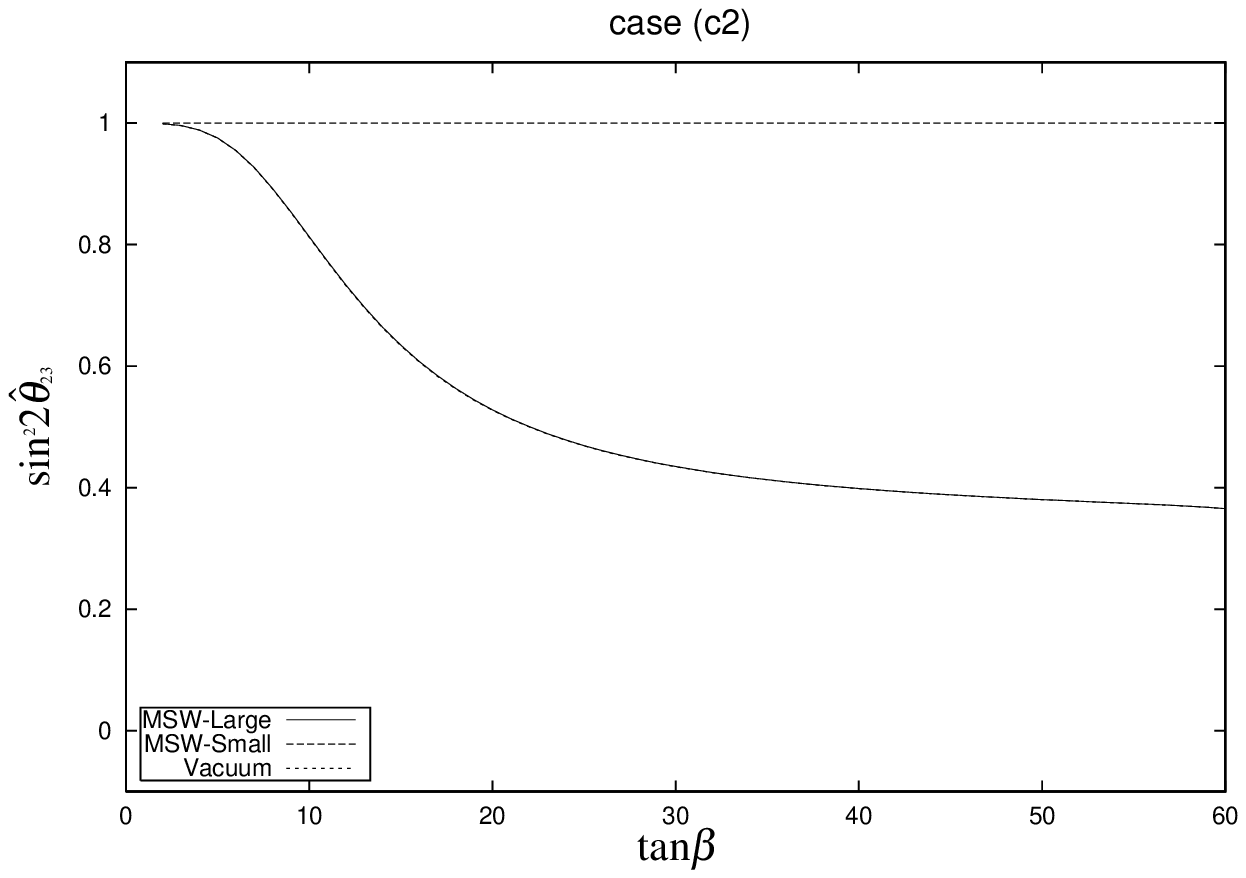}}
 \scalebox{.62}{\includegraphics{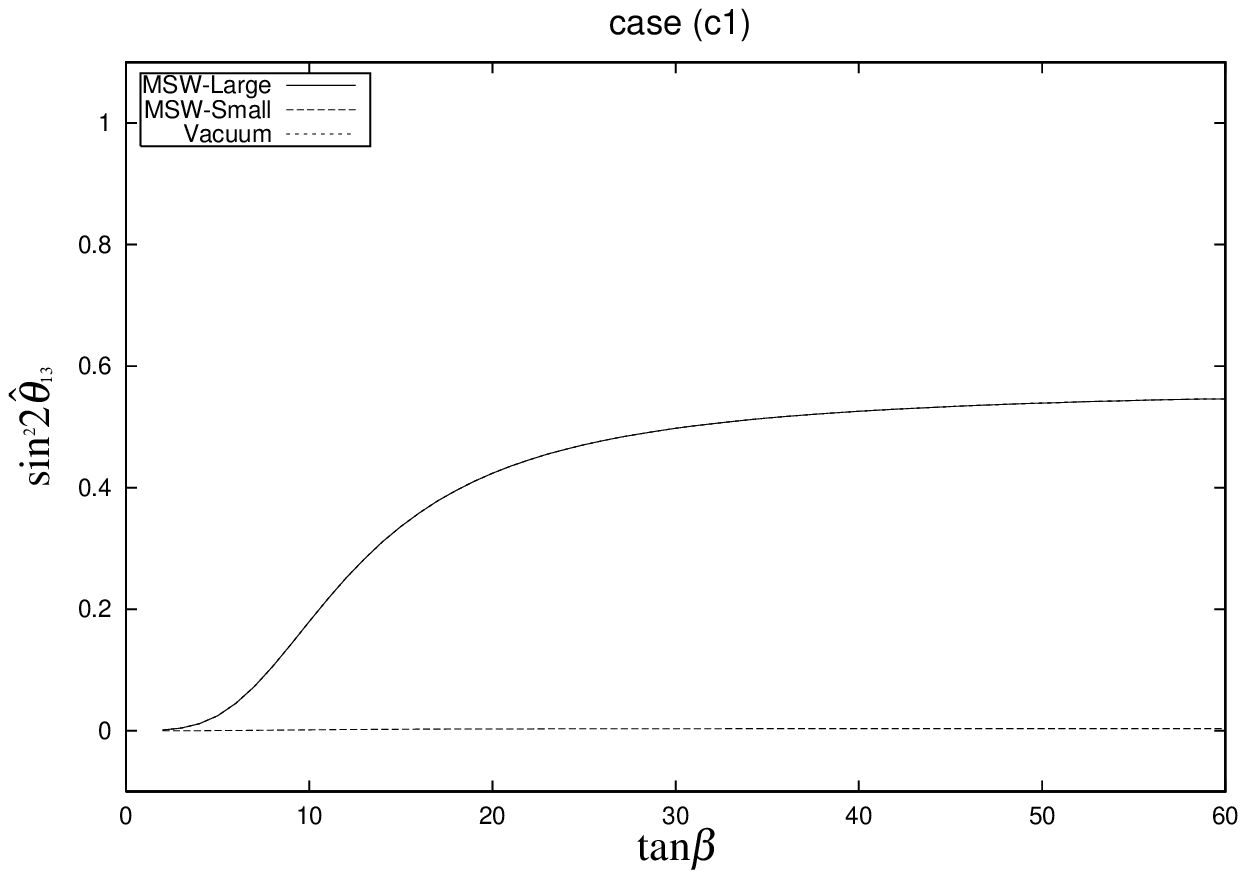}}
 \scalebox{.62}{\includegraphics{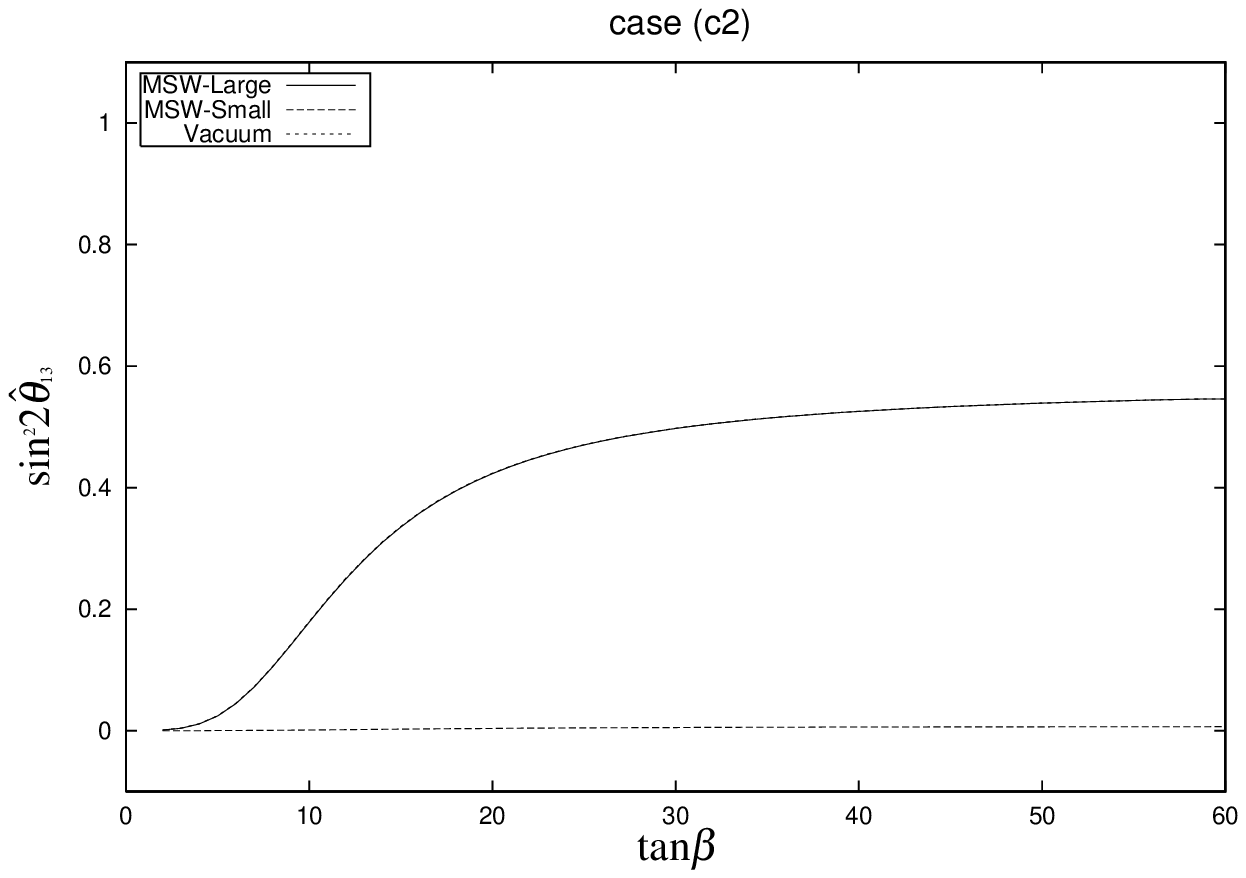}}
 \end{center} 
 \caption{ $\tan \beta$ dependences of $\sin^2 2\hat{\theta}_{ij}$
in (c1) and (c2) with $m_0^{} = 1.0$ eV.}
 \fglab{TypeC1}
\end{figure}
\Fgsref{TypeC1} show the $\tan \beta$ dependences of 
$\sin^2 2 \hat{\theta}_{ij}$ in (c1) and (c2) with $m_0^{}=1.0$ eV.
They shows that all $\sin^2 2\hat{\theta}_{ij}$s 
gradually approach to the fixed values
as $\tan \beta$ becomes large.
We can estimate the value of $\tan \beta$ where the
mixing angles become close to the fixed values.

Dotted-lines of (b) in
\Fgsref{epsilon_e} and \Fgvref{epsilon_mu}
show the value of
\beq
\delta_c^{} =
\dfrac{\Delta m^2_{\rm ATM}}{2m_0^2} = \numt{1.9}{-3}\,,
\mbox{\qquad} (m_0 = 1.0 \mbox{ eV})\,.
\eqlab{c-10}
\eeq
 We can see that $\delta_c^{} \ll \epsilon$
for $20 \leq \tan \beta$ in \Fgsref{epsilon_e} and
\Fgvref{epsilon_mu}.
 As $\tan \beta$ increases,
the quantum effects become larger than 
the effects of mass squared differences of neutrinos.
All $\sin^2 2 \hat{\theta}_{ij}$s approach
to their fixed values around $\tan \beta \sim 20$
as 
are shown in \Fgsref{TypeC1}.

By taking the limit of $\delta_c^{} \ll \epsilon$
we can obtain the fixed values of $\sin^2 2\hat{\theta}_{ij}$
according to the solar neutrino solutions in (c1) and (c2)
as follows.

\bs

\noindent
\UL{case (c1)}:
 The eigenvectors $V_{1,2,3}$ at the $m_z^{}$ scale
are given by
\beq
V_1=
\bma{c}
\cos \theta \\
\dfrac{-1}{\sqrt{2}}\sin \theta  \\
\dfrac{1}{\sqrt{2}} \sin \theta
\ema\,,
\mbox{\quad}
V_2=
\bma{c}
\sin \theta \\
\dfrac{ 1}{\sqrt{2}} \cos \theta  \\
\dfrac{-1}{\sqrt{2}} \cos \theta
\ema\,,
\mbox{\quad}
V_3=
\bma{c}
~~0~~\\
\dfrac{1}{\sqrt{2}} \\
\dfrac{1}{\sqrt{2}}
\ema\,.
\eqlab{OldOnes}
\eeq
 $V_i$ is the eigenvector of the
$i$-th eigenvalue of the neutrino mass matrix,
which corresponds to each column of the MNS matrix in \eqref{MNS2}.
 At the $M_R^{}$ scale,
eigenvalues of \eqref{parts-mr} are given by
\bea
m_1^{}(M_R^{})&=& - m_0^{}(1-\epsilon(1+\cos^2\theta))\,,\nn \\
m_2^{}(M_R^{})&=&m_0^{}(1-2\epsilon)\,, \nn \\
m_3^{}(M_R^{})&=&m_0^{}(1-\epsilon \sin^2 \theta) \,,
\eea
up to the order $\epsilon$
under the condition of $\delta_c^{} \ll \epsilon $.
 Eigenvectors of them are given by
\beq
V_1^{\prime}
=
\bma{c}
\cos \theta \\
\\
-\dfrac{ \sin \theta}{\sqrt{2}} \\
\\
 \dfrac{ \sin \theta}{\sqrt{2}} 
\ema\,,
\mbox{\quad}
V_2^{\prime}
=
\bma{c}
 \dfrac{\sin\theta}{\sqrt{1+\cos^2\theta}} \\
\\
 \dfrac{\sqrt{2}\cos\theta}{\sqrt{1+\cos^2\theta}} \\
\\
0
\ema\,,
\mbox{\quad}
V_3^{\prime}
=
\bma{c}
-\dfrac{1}{2}\dfrac{\sin 2\theta}{\sqrt{1+\cos^2\theta}} \\
\\
 \dfrac{1}{\sqrt{2}}\dfrac{\sin^2\theta}{\sqrt{1+\cos^2\theta}} \\
\\
 \dfrac{1}{\sqrt{2}}{\sqrt{1+\cos^2\theta}} \\
\ema\,.
\eqlab{NewOnesC1}
\eeq
By comparing \eqref{OldOnes} with \eqref{NewOnesC1},
the relation between $V_i^{}$ and $V_i^{\prime}$ is given as
\bea
\bma{c}
V_1^{\prime} \\
\\
V_2^{\prime} \\
\\
V_3^{\prime}
\ema
&=&
\bmaT
~~~~1~~~~ & 0 & 0 \\
\\
0 & \dfrac{1}{\sqrt{1+\cos^2 \theta}} 
    & \dfrac{\cos \theta}{\sqrt{1+\cos^2 \theta}}   \\
\\
0 & \dfrac{-\cos \theta}{\sqrt{1+\cos^2 \theta}}
    & \dfrac{1}{\sqrt{1+\cos^2 \theta}}
\ema
\bma{c}
V_1^{}\\
\\
V_2^{}\\
\\
V_3^{}
\ema\,.
\eqlab{mix-c1}
\eea
When we take $\pi/4$ ($0$) for the parameter $\theta$ in \eqref{mix-c1},
we can obtain the relation between $V_i$ and $V_i^{\prime}$
for the MSW-L and the VO solutions (the MSW-S solution).

For the MSW-L and the VO solutions,
the MNS matrix at the $M_R^{}$ scale
is given by
\beq
\hat{U}_{\rm MNS}=
\bmaT
1/\sqrt{2} & 1/\sqrt{3} & -1/\sqrt{6} \\
-1/2       & 2/\sqrt{3} & 1/2\sqrt{3} \\
 1/2       &   0        & \sqrt{3}/2
\emaT
\eqlab{MNSc1}
\eeq
in the limit of $\delta_c^{} \ll \epsilon$.
The relation between $V_i$ and $V_i^{\prime}$ is given by
\beq
V_1^{\prime}
=
V_1^{}\,,
\mbox{\qquad}
V_2^{\prime}
=
  \sqrt{\dfrac{2}{3}}~ V_2^{}
+ \sqrt{\dfrac{1}{3}}~ V_3^{} \,,
\mbox{\qquad}
V_3^{\prime}
=
-\sqrt{\dfrac{1}{3}}~ V_2^{}
+\sqrt{\dfrac{2}{3}}~ V_3^{}\,.
\eqlab{c1-45}
\eeq

For the MSW-S solution\footnote{{
Although the MSW-S solution in (c1) and (c2) with $m_0^{}=1.0$ eV 
is already excluded by the neutrino-less double-$\beta$ decay
experiments of \eqref{0ne2b-2},
we discuss it here to check whether our 
analytic calculations are consistent with our
numerical results or not.
We will see they are consistent with each other soon.
}},
the MNS matrix at the $M_R^{}$ is given by
\beq
\hat{U}_{\rm MNS}=
\bmaT
1 & 0 & 0 \\
0 & 1 & 0 \\
0 & 0 & 1
\emaT\,,
\eqlab{MNSc1-0}
\eeq
which means
\beq
V_1^{\prime}
=
V_1^{}\,,
\mbox{\qquad}
V_2^{\prime}
= \dfrac{1}{\sqrt{2}}~V_2^{} + \dfrac{1}{\sqrt{2}}~V_3^{} \,,
\mbox{\qquad}
V_3^{\prime}
= -\dfrac{1}{\sqrt{2}}~V_2^{} + \dfrac{1}{\sqrt{2}}~V_3^{}\,.
\eqlab{c1-0}
\eeq

\bs

\noindent
\UL{case (c2)}:
By the same calculations as those of (c1),
eigenvalues of \eqref{parts-mr} are obtained as
\bea
m_1^{}(M_R^{})&=&m_0^{}(1-2\epsilon)\,, \nn \\
m_2^{}(M_R^{})&=&-m_0^{}(1-\epsilon(1+\sin^2\theta))\,,\nn \\
m_3^{}(M_R^{})&=&m_0^{}(1-\epsilon \cos^2 \theta) \,,
\eea
up to the $O(\epsilon)$ in $\delta_c^{}\ll\epsilon$.
Eigenvectors of them are given by
\beq
V_1^{\prime}
=
\bma{c}
\dfrac{\cos \theta}{\sqrt{1+\sin^2 \theta}} \\
\\
-\dfrac{\sqrt{2}\sin \theta}{\sqrt{1+\sin^2 \theta}} \\
\\
0
\ema\,,
\mbox{\quad}
V_2^{\prime}
=
\bma{c}
\sin \theta\\
\\
\dfrac{\cos \theta}{\sqrt{2}} \\
\\
-\dfrac{\cos \theta}{\sqrt{2}}
\ema\,,
\mbox{\quad}
V_3^{\prime}
=
\bma{c}
\dfrac{1}{2}\dfrac{\sin 2 \theta}{\sqrt{1+\sin^2\theta}} \\
\\
\dfrac{1}{\sqrt{2}}\dfrac{\cos^2\theta}{\sqrt{1+\sin^2\theta}} \\
\\
\dfrac{1}{\sqrt{2}}{\sqrt{1+\sin^2\theta}} \\
\ema\,.
\eqlab{NewC2}
\eeq
By comparing \eqref{OldOnes} with \eqref{NewC2},
the relation between $V_i$ and $V_i^{\prime}$ is given by 
\bea
\bma{c}
V_1^{\prime} \\
\\
V_2^{\prime} \\
\\
V_3^{\prime}
\ema
&=&
\bmaT
\dfrac{1}{\sqrt{1+\sin^2 \theta}} 
&~~~~ 0~~~~
& \dfrac{-\sin \theta}{\sqrt{1+\sin^2 \theta}}   \\
\\
0 &~~~~ 1~~~~ &0 \\
\\
\dfrac{\sin \theta}{\sqrt{1+\sin^2 \theta}}
&~~~~ 0~~~~
    & \dfrac{1}{\sqrt{1+\sin^2 \theta}}
\ema
\bma{c}
V_1^{} \\
\\
V_2^{} \\
\\
V_3^{}
\ema\,.
\eqlab{mix-c2}
\eea

For the MSW-L and the VO solutions,
the MNS matrix at the $M_R^{}$ scale
is given by 
\beq
\hat{U}_{\rm MNS}=
\bmaT
1/\sqrt{3} & 1/\sqrt{2} & 1/\sqrt{6} \\
-\sqrt{2/3}& 1/2        & 1/2\sqrt{3} \\
 0         & -1/2       & \sqrt{3}/2
\emaT
\eqlab{MNSc2}
\eeq
in $\delta_c^{} \ll \epsilon$.
This means
\beq
V_2^{\prime} = V_2^{}\,,
\mbox{\qquad}
V_1^{\prime}=
\sqrt{\dfrac{2}{3}}~ V_1^{} -\sqrt{\dfrac{1}{3}}~V_3^{}\,,
\mbox{\qquad}
V_3^{\prime}=
\sqrt{\dfrac{1}{3}}~ V_1^{} + \sqrt{\dfrac{2}{3}}~V_3^{}\,.
\eqlab{c2-45}
\eeq
This is consistent with the results in Ref.~\cite{EllisLola}.

For the MSW-S solution$^{\|}$,
the MNS matrix at the $M_R^{}$ scale is obtained as
\beq
\hat{U}_{\rm MNS}=
\bmaT
1 & 0 & 0 \\
0 & 1/\sqrt{2} & 1/\sqrt{2} \\
0 & -1\sqrt{2} & 1/\sqrt{2}
\emaT\,,
\eqlab{MNSc2-0}
\eeq
which suggests
\beq
V_1^{\prime} = V_1^{}\,,
\mbox{\qquad}
V_2^{\prime} = V_2^{}\,,
\mbox{\qquad}
V_3^{\prime} = V_3^{}\,.
\eqlab{c2-0}
\eeq

\bs

For the MSW-L and the VO solutions in (c1) and (c2),
\eqsref{MNSc1} and \eqvref{MNSc2} suggest
the fixed values of the $\sin^2 2\hat{\theta}_{ij}$s are
\beq
\sin^2 2\hat{\theta}_{12} = 0.96\,,
\mbox{\quad}
\sin^2 2\hat{\theta}_{23} = 0.36\,,
\mbox{\quad}
\sin^2 2\hat{\theta}_{13} = \dfrac{5}{9}\,,
\eqlab{fix1}
\eeq
in the limit of $\epsilon \gg \delta_c^{}$.
We cannot see the differences between the MSW-L and
the VO solutions in \Fgsref{TypeC1}.
It is because
the value of $\tan \beta$ where all $\sin^2 2\hat{\theta}_{ij}$s
approach to their fixed values are determined
by $\epsilon$ and $\delta_c^{}$ which
does not  relate to  $\Delta m^2_{\rm solar}$
but to $\Delta m^2_{\rm ATM}$.
 The rearrangements are occurred between
$V_2$ and $V_3$ in (c1), and $V_1$ and $V_3$ in (c2),
where the squared mass differences of their eigen values
are mainly determined by $\Delta m^2_{\rm ATM}$.
On the other hand,
for the MSW-S solution
\eqsref{MNSc1-0} and \eqvref{MNSc2-0} suggest
all mixing angles approach to zero in the case of (c1),
and 
$\sin^2 2\hat{\theta}_{12}=\sin^2 2\hat{\theta}_{13}=0$,
$\sin^2 2\hat{\theta}_{23}=1$ in the case of (c2).
These results from analytic calculations are completely consistent with
those from the numerical analyses as shown in \Fgref{TypeC1}.

\subsubsection{$m_0=0.2$ eV}
\Fgsref{TypeC2-1}, \Fgvref{TypeC2-2},
\Fgvref{TypeC2-3} and \Fgvref{TypeC2-4}
show the $\tan \beta$ dependences of $\sin^2 2\hat{\theta}_{ij}^{}$s
with $m_0=0.2$ eV in 
(c1), (c2), (c3) and (c4), respectively.
\begin{figure}[t]
 \scalebox{.62}{\includegraphics{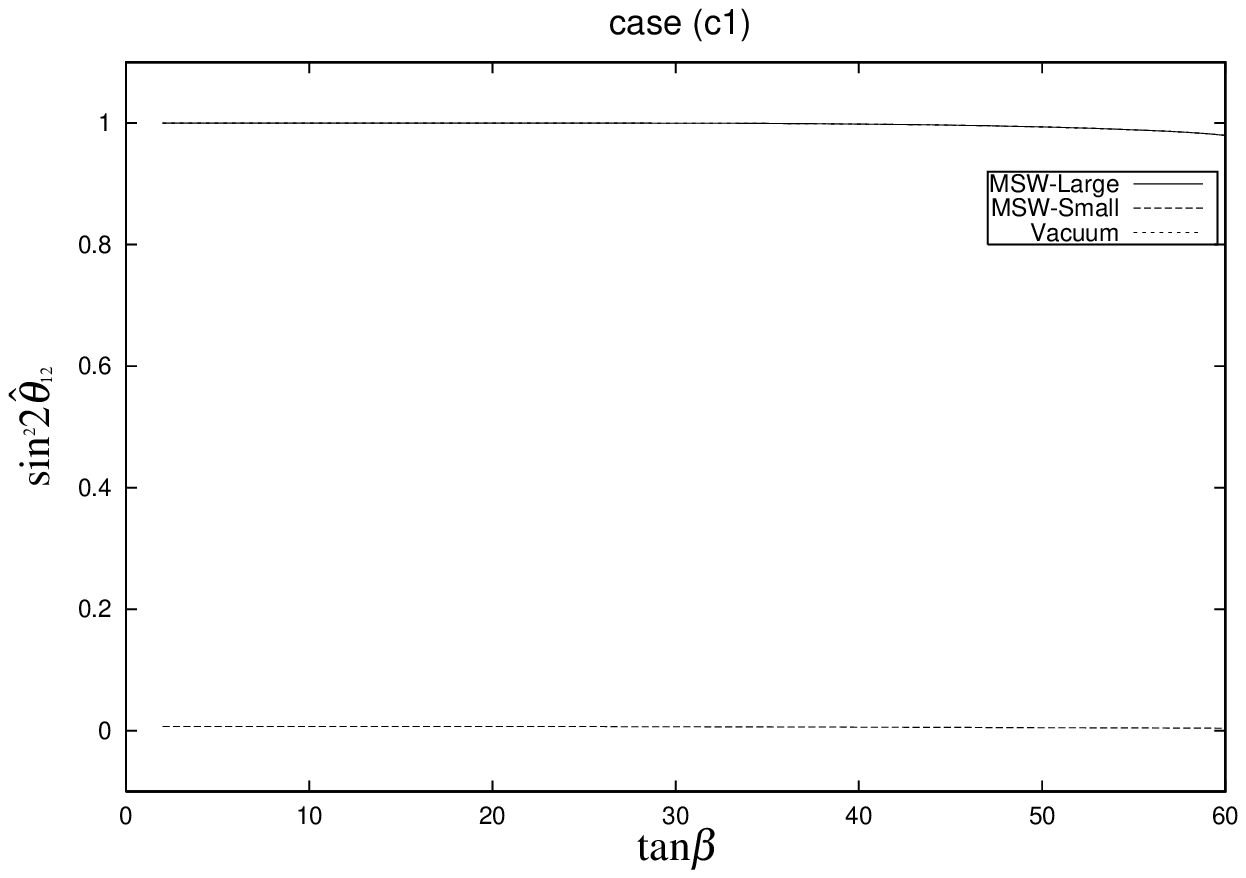}}
 \scalebox{.62}{\includegraphics{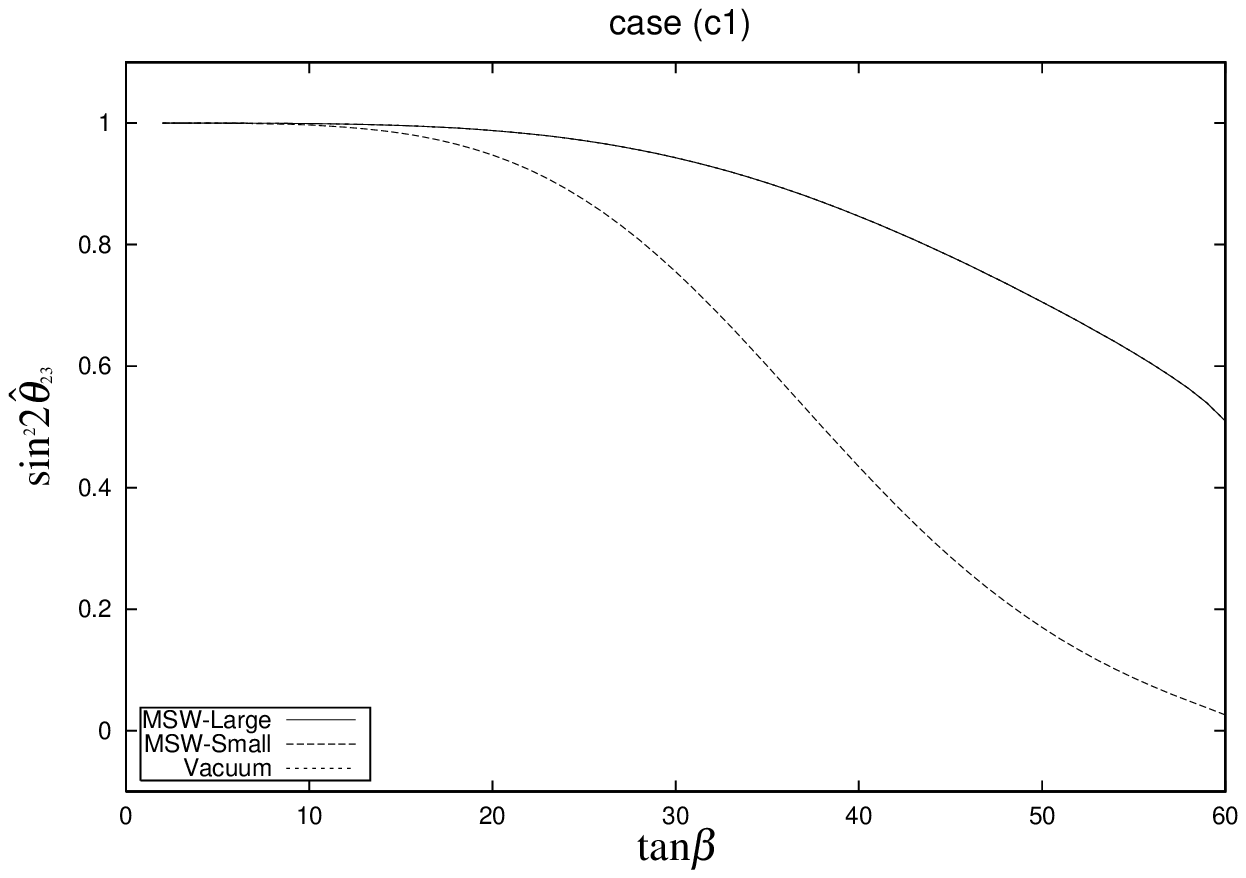}}
 \scalebox{.62}{\includegraphics{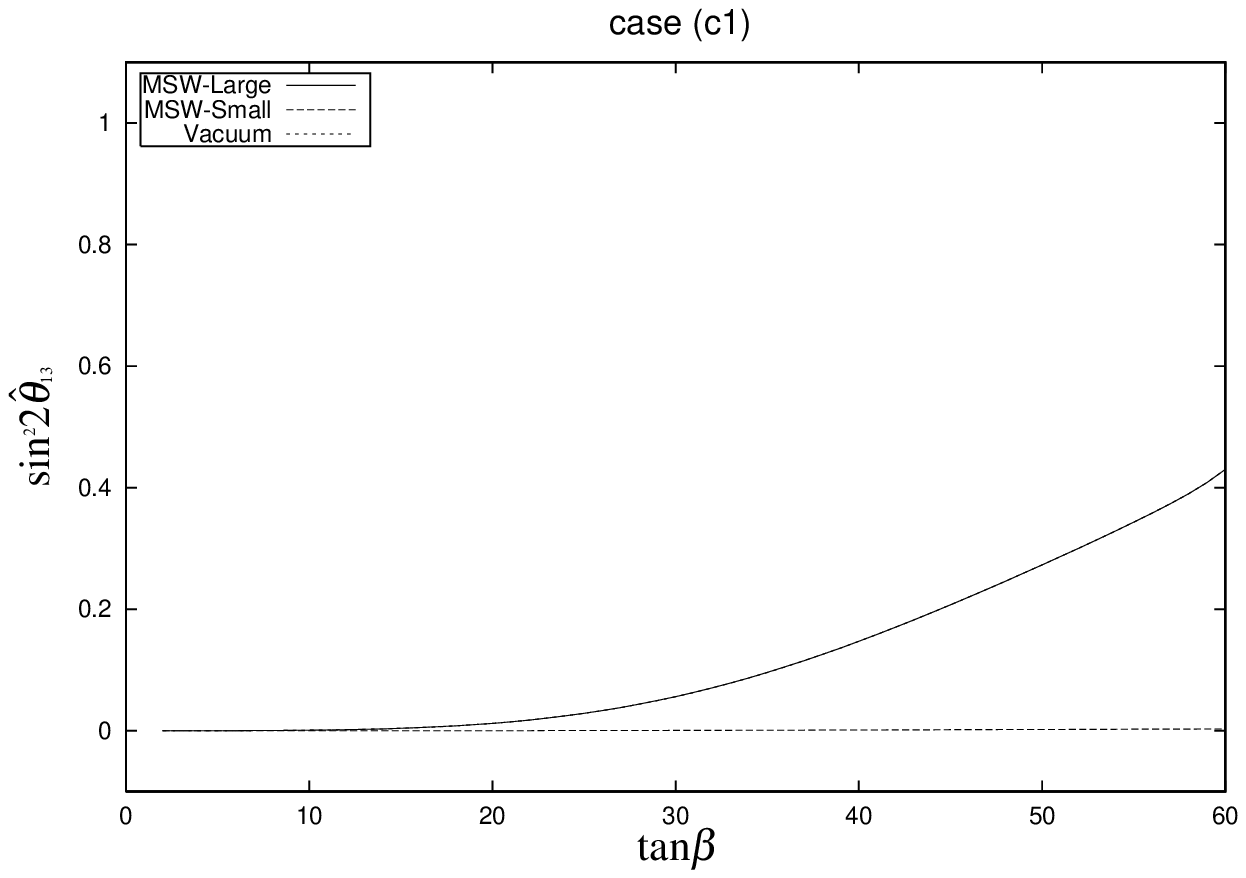}}
 \caption{ $\tan \beta$ dependences of $\sin^2 2 \hat{\theta}_{ij}$
in (c1) with $m_0^{}=0.2$ eV.}
 \fglab{TypeC2-1}
\end{figure}
\begin{figure}[t]
 \scalebox{.62}{\includegraphics{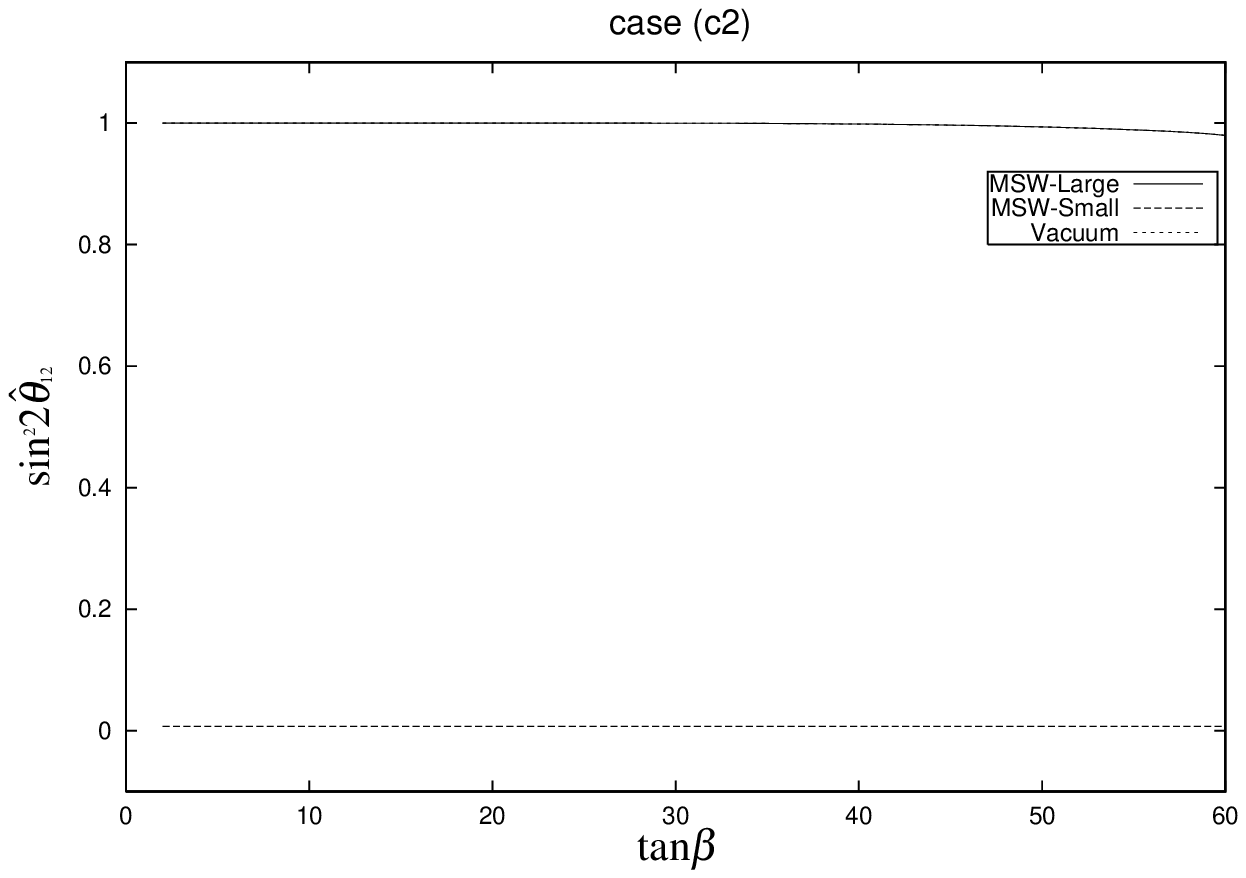}}
 \scalebox{.62}{\includegraphics{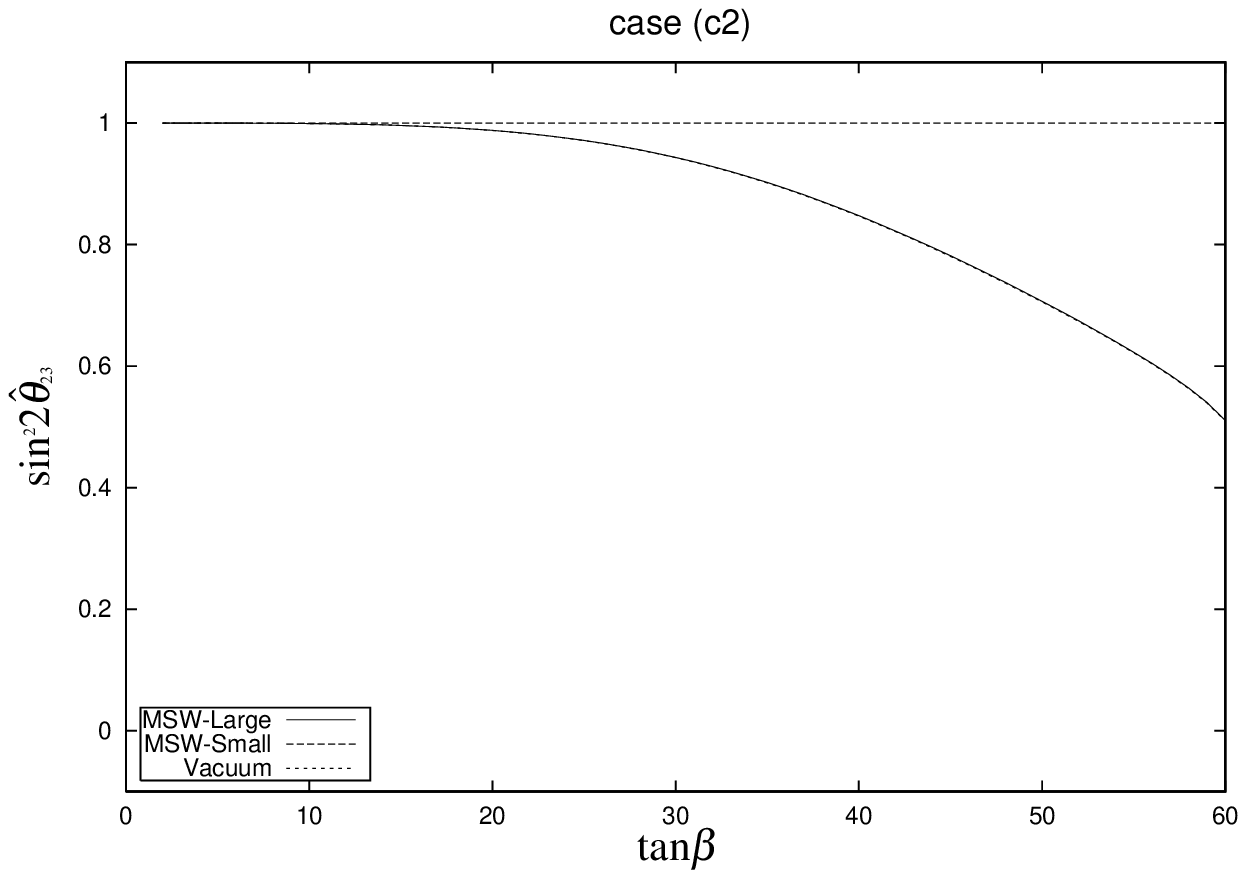}}
 \scalebox{.62}{\includegraphics{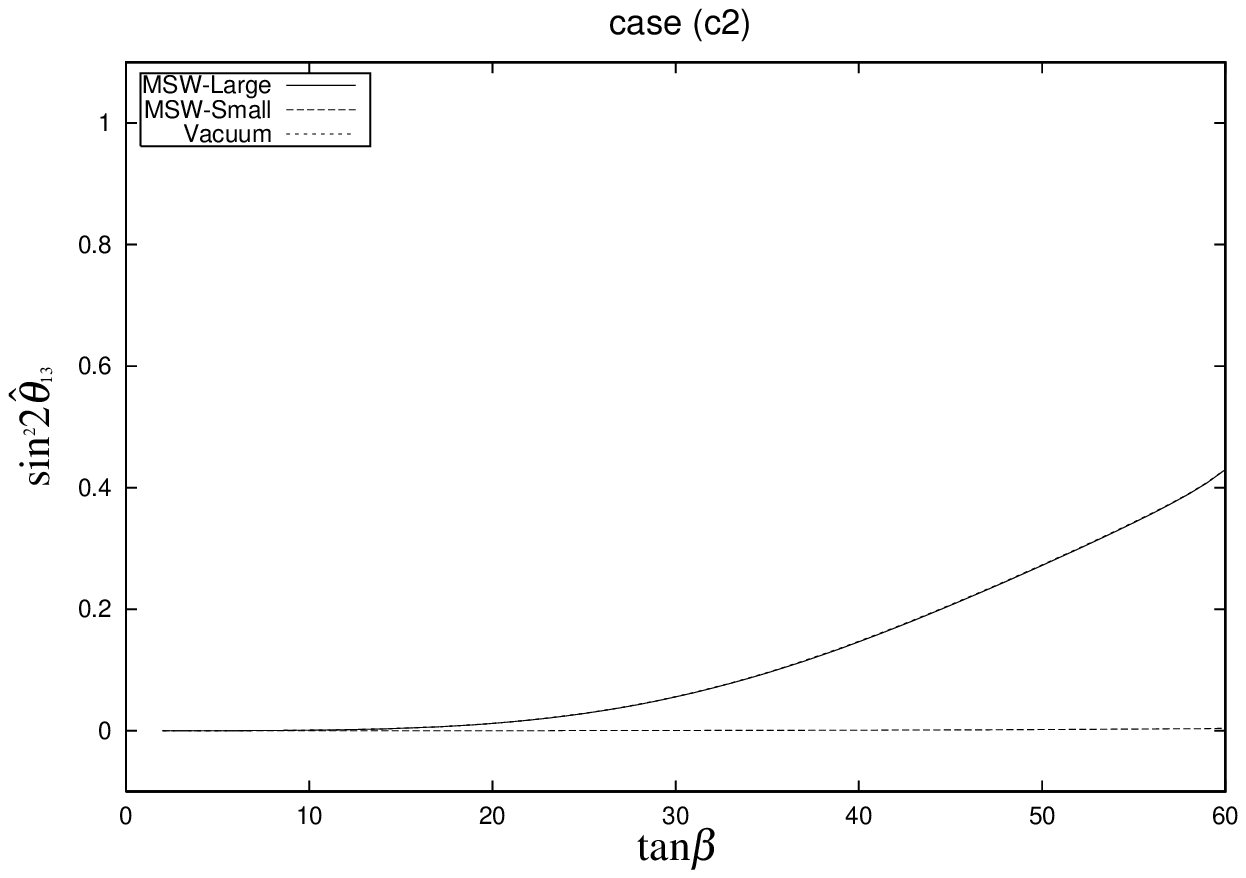}}
 \caption{ $\tan \beta$ dependences of $\sin^2 2 \hat{\theta}_{ij}$
in (c2) with $m_0^{}=0.2$ eV.}
 \fglab{TypeC2-2}
\end{figure}
\begin{figure}[t]
 \scalebox{.62}{\includegraphics{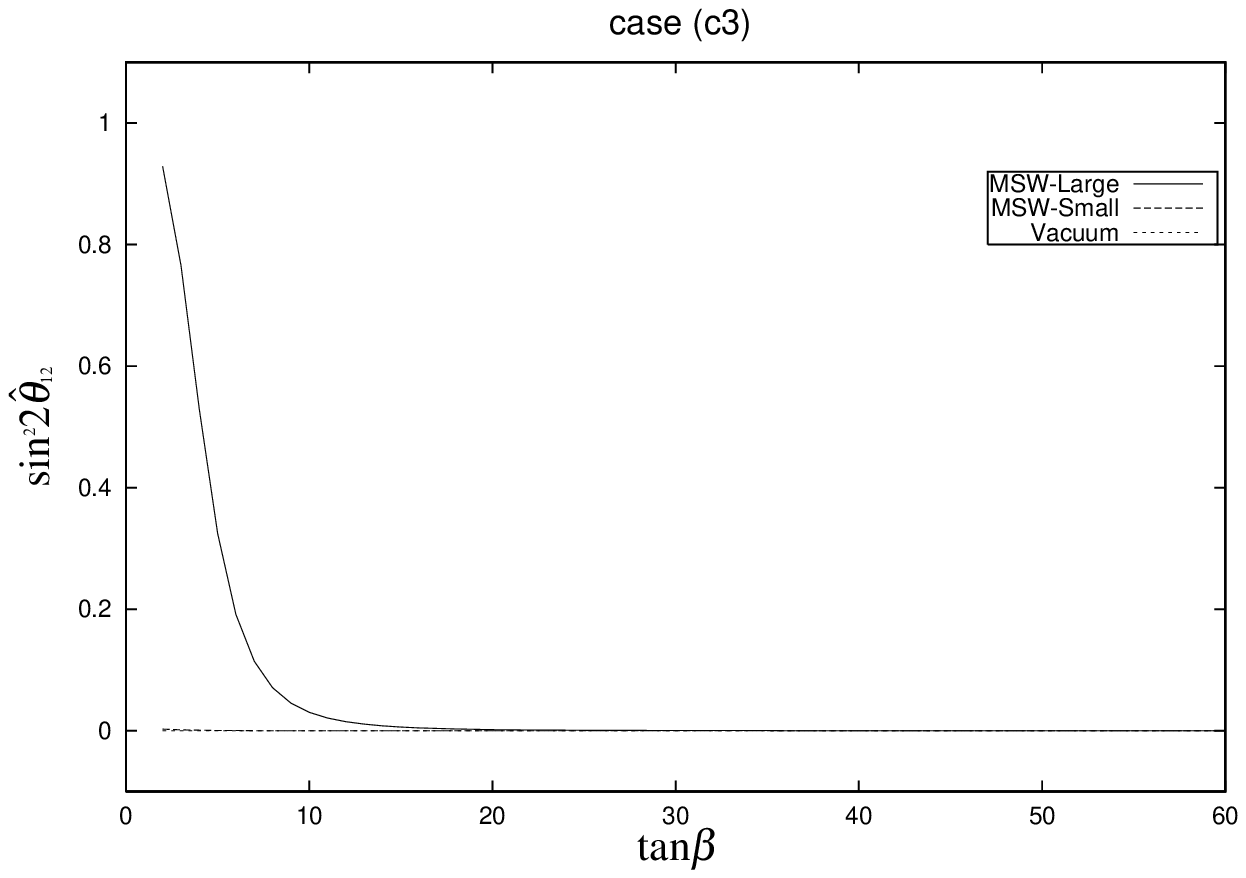}}
 \scalebox{.62}{\includegraphics{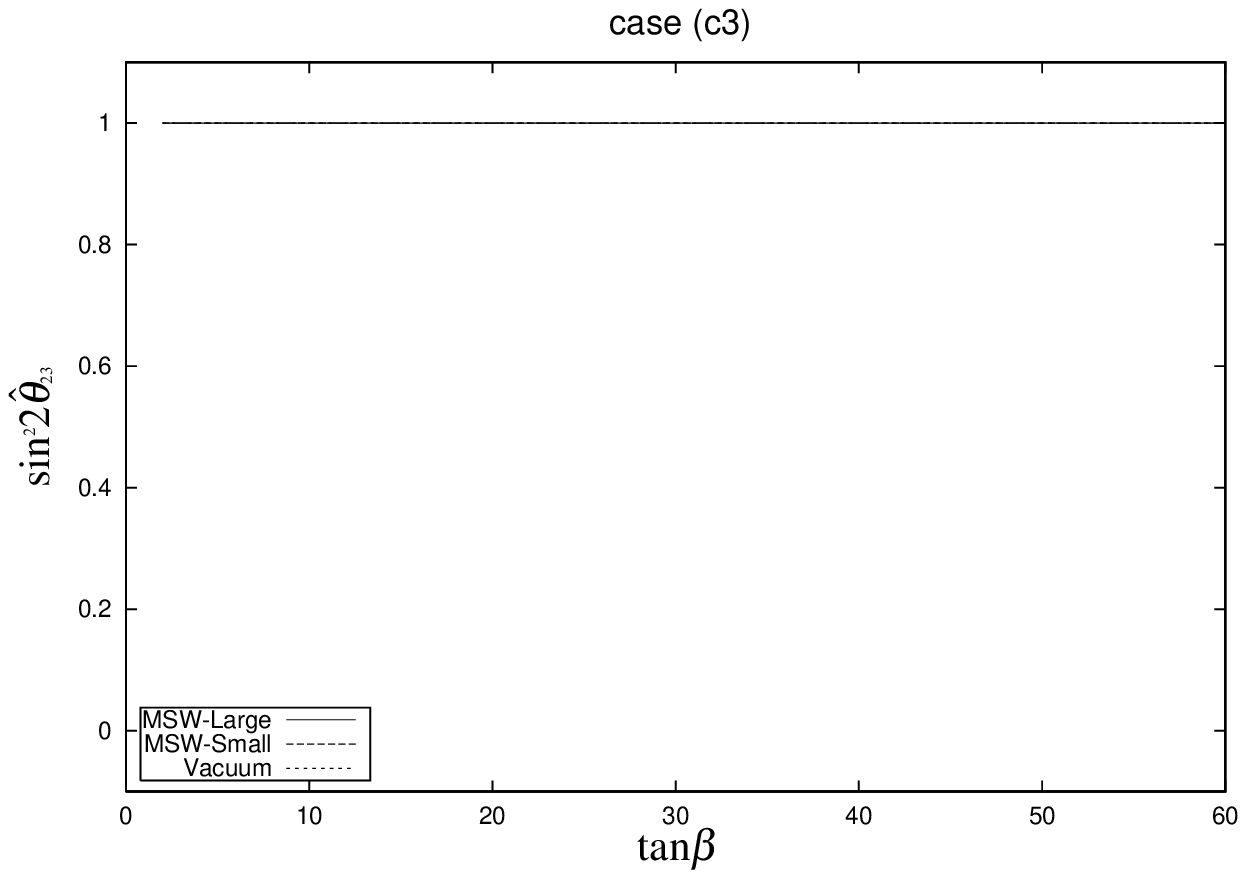}}
 \scalebox{.62}{\includegraphics{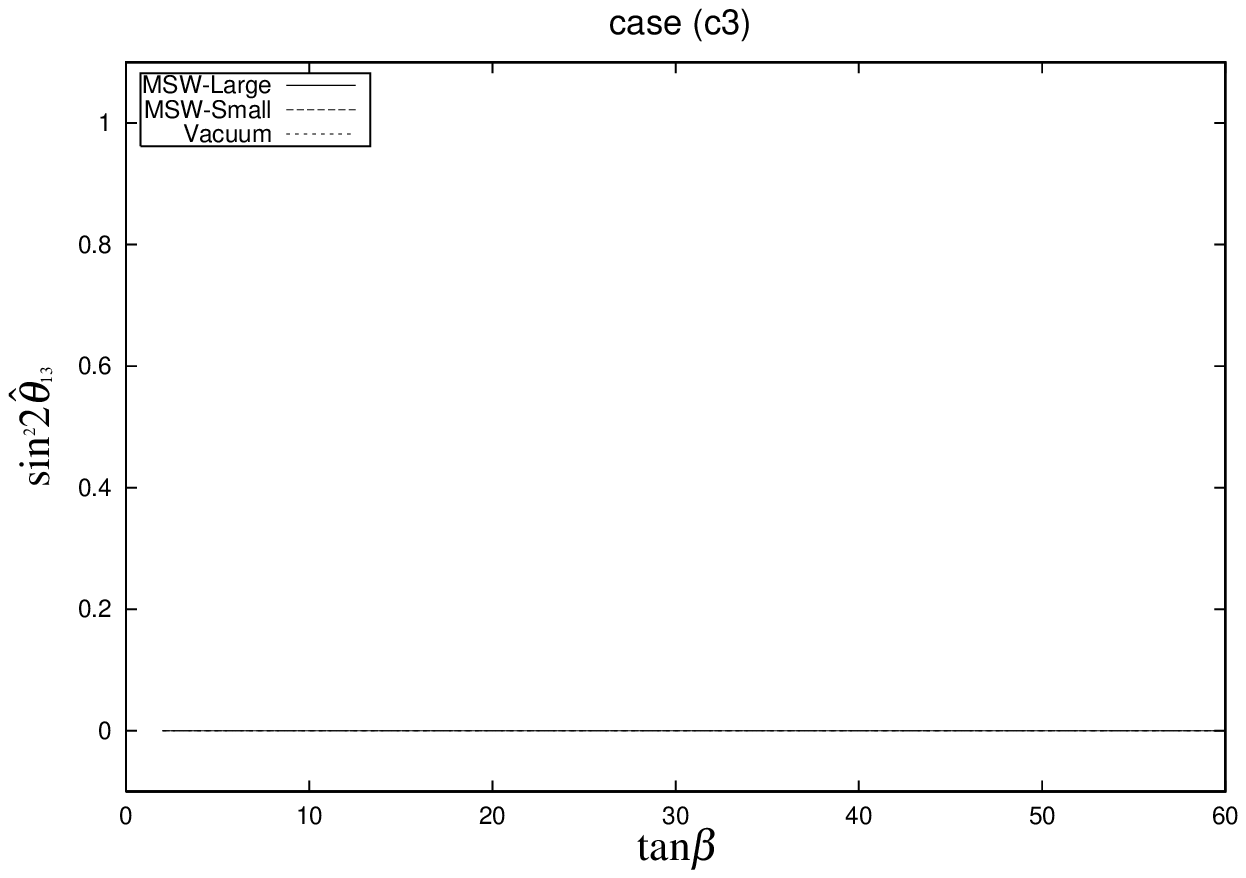}}
 \caption{ $\tan \beta$ dependences of $\sin^2 2 \hat{\theta}_{ij}$
in (c3) with $m_0^{}=0.2$ eV.}
 \fglab{TypeC2-3}
\end{figure}
\begin{figure}[t]
 \scalebox{.62}{\includegraphics{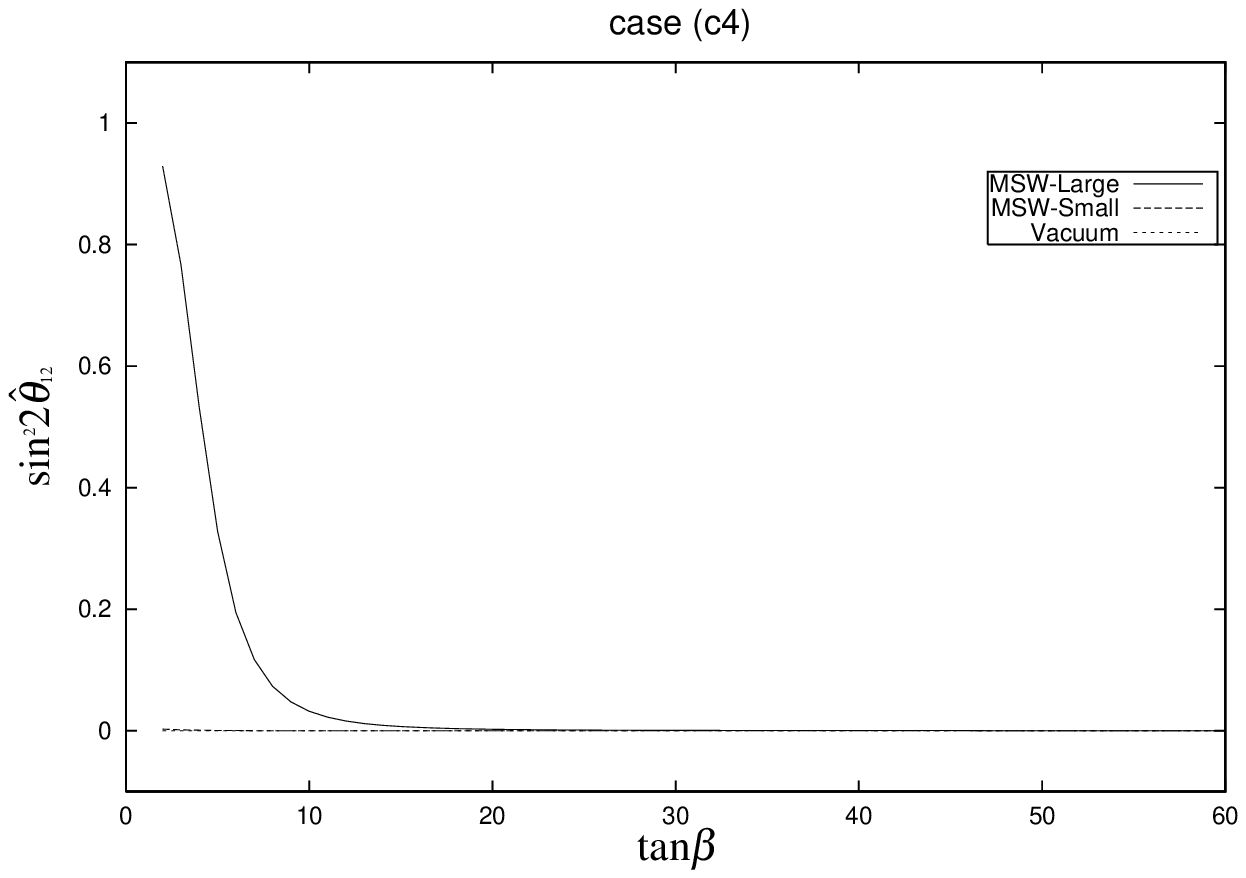}}
 \scalebox{.62}{\includegraphics{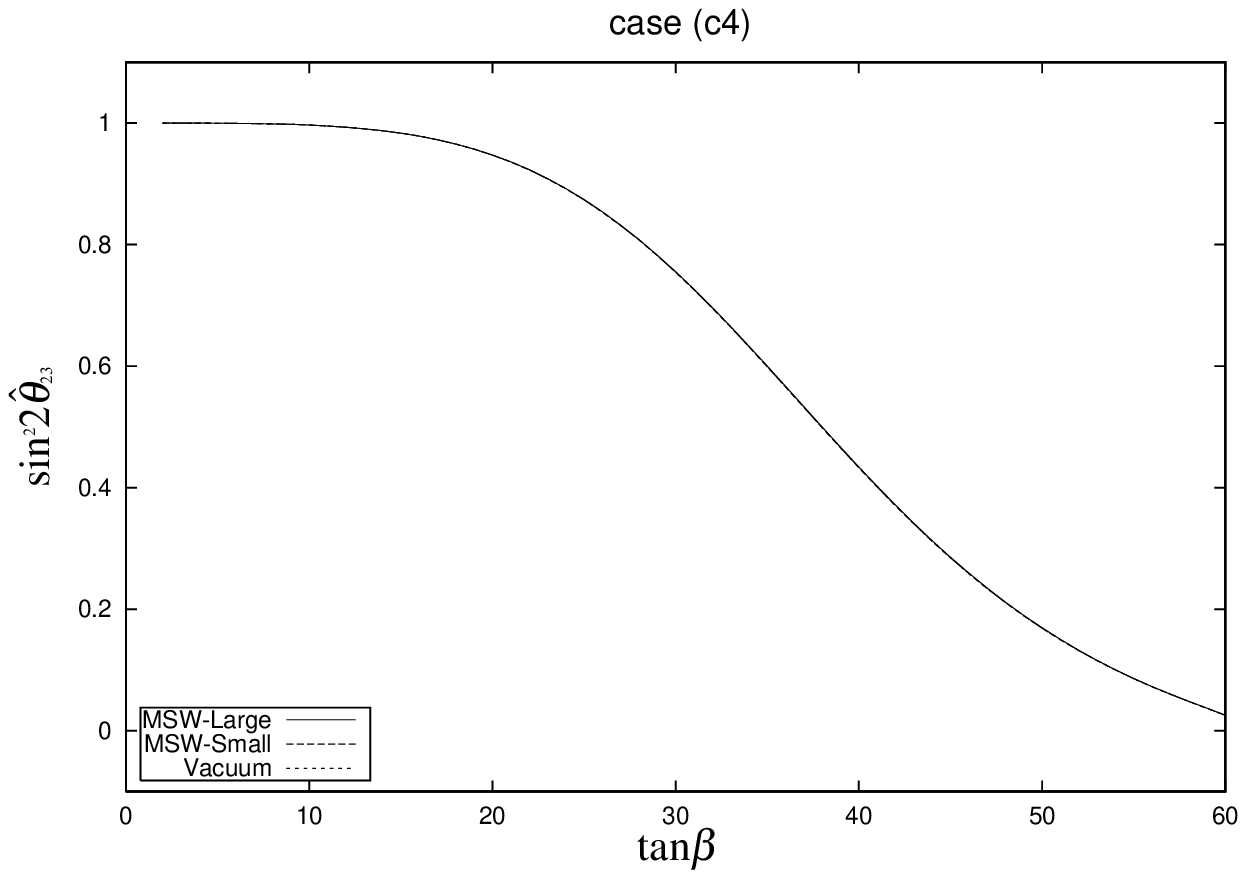}}
 \scalebox{.62}{\includegraphics{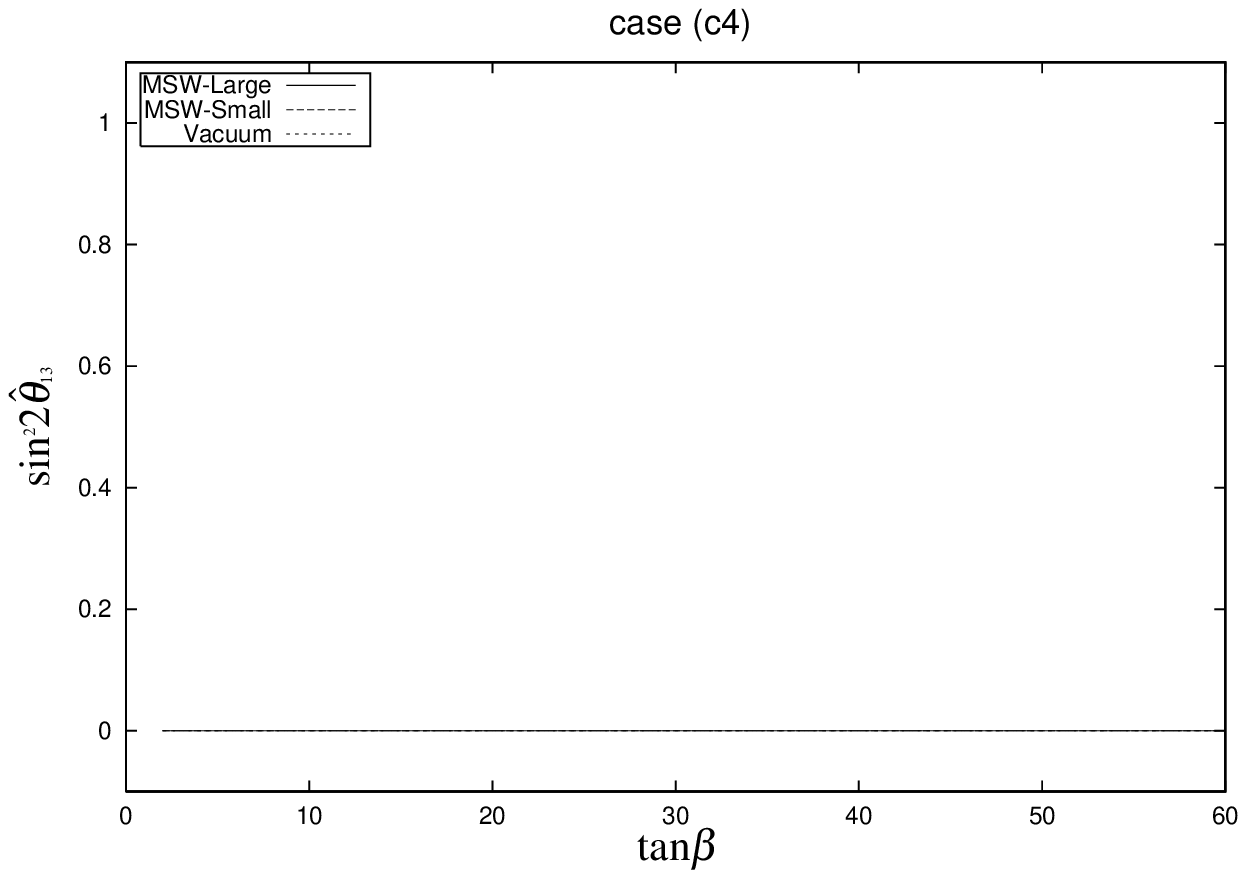}}
 \caption{ $\tan \beta$ dependences of $\sin^2 2 \hat{\theta}_{ij}$
in (c4) with $m_0^{}=0.2$ eV.}
 \fglab{TypeC2-4}
\end{figure}

\bs

\noindent
\UL{cases (c1) and (c2)}:
 Figures~\ref{fig:TypeC2-1} and \ref{fig:TypeC2-2} show that
all $\sin^2 2\hat{\theta}_{ij}$s approach to the same fixed values
as those with $m_0^{}=1.0$ eV
in the large $\tan \beta$ region.
However, the value of $\tan \beta$
where $\sin^2 2\hat{\theta}_{ij}$s becomes
close to their fixed values with $m_0^{}=0.2$ eV is larger than 
that with $m_0^{}=1.0$ eV.
 The value of $\delta_c^{}$ with $m_0^{}=0.2$ eV is given by
\beq
\delta_c^{} =
\dfrac{\Delta m^2_{\rm ATM}}{2m_0^2} = \numt{4.6}{-2}\,,
\mbox{\qquad}
(m_0^{}=0.2 \mbox{ eV})\,,
\eqlab{unitMass-ATM}\\
\eeq
which is shown as dotted-lines of
(a) in \Fgsref{epsilon_e} and \Fgvref{epsilon_mu}.
 In the region of $2\leq\tan \beta < 50$,
$\epsilon$ is not larger than $\delta_c^{}$, and
the first (second) generation cannot be regarded to be
degenerate with the third generation.
 Thus, the rearrangement 
between the eigenvectors of $V_1(V_2)$ and $V_3$
is not realized completely
although the sign of $m_1^{}(m_2^{})$ is the same as that of
$m_3^{}$.
On the other hands,
the rearrangement between these eigenvectors 
is realized with $m_0^{}=1.0$ eV in the region of $\tan \beta \geq 20$,
where $\epsilon$ is already larger than $\delta_c^{}$.

\bs

\noindent
\UL{case (c3)}:
 \Fgsref{TypeC2-3} show that 
for all solar neutrino solutions
$\sin^2 2\hat{\theta}_{12}$ and
$\sin^2 2\hat{\theta}_{13}$
are almost zero in the region of $\tan \beta \geq 10$, and
$\sin^2 2\hat{\theta}_{23} \simeq 1$ in all $\tan \beta$ region.
Up to the $O(\epsilon)$
eigenvalues of \eqref{parts-mr} are given by
\bea
m_1^{}(M_R^{})&=&-m_0^{}(1-2\epsilon)\,,\nn \\
m_2^{}(M_R^{})&=&-m_0^{}(1- \epsilon)\,, \nn \\
m_3^{}(M_R^{})&=& m_0^{}(1- \epsilon) \,,
\eea
and eigenvectors of them are given by
\beq
V_1^{\prime}=
\bma{c}
1 \\
0 \\
0
\ema,
\mbox{\quad}
V_2^{\prime}=
\bma{c}
0 \\
{1}/{\sqrt{2}} \\
{-1}/{\sqrt{2}}
\ema,
\mbox{\quad}
V_3^{\prime}=
\bma{c}
0 \\
{1}/{\sqrt{2}} \\
{1}/{\sqrt{2}}
\ema\,.
\eqlab{NewOnesC3}
\eeq
 These eigenvectors and eigenvalues 
do not depend on the mixing angle $\theta$.
\Eqsref{NewOnesC3} suggest
\beq
\sin^2 2\hat{\theta}_{12} = 0\,,
\mbox{\qquad} 
\sin^2 2\hat{\theta}_{23} = 1\,,
\mbox{\qquad} 
\sin^2 2\hat{\theta}_{13} = 0\,,
\mbox{\qquad} 
\eeq
from \eqsref{sin2_2theta} in the region of $\xi_c^{} \ll \epsilon$.
By comparing \eqref{OldOnes} with \eqref{NewOnesC3},
we can obtain 
\bea
\bma{c}
V_1^{\prime} \\
V_2^{\prime} \\
V_3^{\prime}
\ema
&=&
\bmaT
 \cos \theta & \sin \theta &0 \\
-\sin \theta & \cos \theta &0 \\
        0    &           0 &~~~1~~~ \\
\ema
\bma{c}
V_1^{} \\
V_2^{} \\
V_3^{}
\ema\,.
\eqlab{mix-c3}
\eea
The rearrangement between $V_1$ and $V_2$
is realized, because the sign of $m_1^{}$ is the 
same as that of $m_2^{}$.

\Fgsref{TypeC2-3} show
 $\sin^2 2\theta_{13}$ and $\sin^2 2\theta_{23}$
are not changed against quantum corrections,
and
$\sin^2 2\hat{\theta}_{12}$ is close to zero
in $\tan \beta \geq 10$ for the MSW-L solution, 
while $\sin^2 2\hat{\theta}_{12} \simeq 0$ 
in all $\tan \beta$ region for the VO solution.
These situations can be explained by 
estimating the value of $\tan \beta$
where the mixing angles are close to the fixed values.
 For the MSW-L solution,
the value of $\xi_c^{}$ is given by
\beq
\xi_c^{} =
\dfrac{\Delta m^2_{\rm MSW-L}}{2m_0^2} = \numt{2.3}{-4}\,.
\eqlab{unitMass-solar}
\eeq
which is shown as dotted-lines of (d)
in Figures~\ref{fig:epsilon_e} and \ref{fig:epsilon_mu}.
Since
$\xi_c^{}$ is much smaller than $\epsilon$
in the region of $10 \leq \tan \beta$,
the first and the second generations are 
regarded to be degenerate at $m_0^{}$,
and the rearrangement between $V_1$ and $V_2$
is realized.
 On the other hand, the VO solution gives
the value of $\xi_c^{}$ as
\beq
 \xi_c^{} =
\dfrac{\Delta m^2_{\rm VO}}{2m_0^2} = \numt{1.1}{-9}\,,
\eqlab{unitMass-VO}
\eeq
which is much smaller than $\epsilon$ in all $\tan \beta$ region.
Therefor the rearrangement between $V_1$ and $V_2$
is realized even in the small $\tan \beta$ region for the VO solution. 

\bs

\noindent
\UL{case (c4)}:
 \Fgsref{TypeC2-4} show all $\sin^2 2\hat{\theta}_{ij}$s approach 
to zero in the large $\tan \beta$ region. 
This means 
 the MNS matrix becomes the 
unit matrix in the limit of $\epsilon \gg \delta_c^{}$.
 In the case of (c4) we cannot obtain the rearrangement
rule between $V_i^{}$ and $V_i^{\prime}$,
because $\kappa$ becomes proportional to the unit matrix
which can be diagonalized by any unitary matrices.

Let us see the MSW-L solution at first.
Figures \ref{fig:epsilon_e} and \ref{fig:epsilon_mu} show 
$\xi_c^{}\ll \delta_c^{} \sim \epsilon$
in the region of $10 < \tan \beta < 50$,
where the value of $\xi_c^{}$ ($\delta_c^{}$) is shown in 
\eqref{unitMass-solar} (\eqref{unitMass-ATM}).
In this region \eqref{parts-mr} becomes
\beq
M_\nu^{}(M_R^{})\simeq m_0^{}
\bmaT
1-2 \epsilon &           0       & 0 \\
0            & \l(1-2\epsilon\r) & \l(1-\epsilon\r) \delta_c^{}/2 \\
0            &\l(1-\epsilon\r) \delta_c^{}/2 & 1
\emaT\,.
\eqlab{c4-mass}
\eeq
This means
$\sin^2 2\hat{\theta}_{12}$ approaches to zero
in $10 < \tan \beta$.
\Eqref{c4-mass} also suggests
the mixing 
between the second and the third generations as
\beq
\tan 2 \hat{\theta}_{23} \simeq
\dfrac{\delta_c^{}}{2\epsilon}\,.
\eqlab{c4-23}
\eeq
In $\tan \beta > 50$,
$\epsilon$ is larger than $\delta_c^{}$, and
$\sin^2 2\hat{\theta}_{23}$ becomes small
as we can see in \Fgsref{TypeC2-4}.

For the VO solution,
since the value of $\xi_c^{}$
in \eqref{unitMass-VO}
is much smaller than values of $\epsilon$ and $\delta_c^{}$,
$M_\nu(M_R^{})$ also becomes \eqref{c4-mass}
in all $\tan \beta$ region.
 Therefor, $\sin^2 2\hat{\theta}_{12}$ and $\sin^2 2\hat{\theta}_{13}$
are zero at any value of $\tan \beta$.
 The behavior of $\sin^2 2\hat{\theta}_{23}$ in the VO solution 
is the same as that in the MSW-L solution,
since \eqref{c4-mass} is independent of $\xi_c^{}$.

Since the mixing angle $\theta\simeq0$
for the MSW-S solution,
$M_\nu(M_R^{})$ also becomes \eqref{c4-mass}.
 Thus $\sin^2 2\hat{\theta}_{12}$ and $\sin^2 2\hat{\theta}_{13}$
are zero in all $\tan \beta$ region, and
$\tan \beta$ dependence of 
$\sin^2 2\hat{\theta}_{23}$ is the same as
that in the MSW-L (VO) solution.

\section{Conclusion}
\label{sec:LAST}
\setcounter{equation}{0}

In this article,
 we study the stability of the Maki-Nakagawa-Sakata (MNS)
lepton-flavor mixing matrix
 against quantum corrections
 in the minimal supersymmetric Standard Model (MSSM)
 with effective dimension-five operators
 which give Majorana masses of neutrinos.
 We decide parameters of the MNS matrix
at the weak scale from the data of experiments,
and obtain the MNS matrix at the high-energy scale
by calculating the quantum corrections.
 Then we analyze the stability of the MNS matrix at the high energy
scale according to types of neutrino mass hierarchy.

\bs

 In the two generation neutrinos,
the mixing angles of Type A$^{(2)}$ ($\kappa_{3}\gg\kappa_{2}$)
and Type B2$^{(2)}$ ($\kappa_{3}\simeq-\kappa_{2}$) are stable against 
quantum corrections,
where $\kappa_{i}$s are eigenvalues of 
$2 \times 2$ neutrino mass matrix at the weak scale.
 The mixing angle of Type B1$^{(2)}$ ($\kappa_{3}\simeq\kappa_{2}$)
is unstable around $\theta_{23}=\pi/4$
when the magnitude of 
the quantum correction $\epsilon$ is larger than that of $\delta k/\kappa_3$.

\bs

In the three generation neutrinos, the stability of the MNS
matrix strongly depends on the types of mass hierarchy and
the relative sign assignments of mass eigenvalues.
The results are obtained as follows.

\begin{enumerate}
\item 
Type A ($m_1^{} \sim m_2^{} \ll m_3^{}$)

The MNS matrix is stable against quantum corrections.

\item
Type B ($m_1^{}\sim m_2^{}\gg m_3^{}$)

$\sin^2 2 {\theta}_{13}$ and $\sin^2 2 {\theta}_{23}$
are stable against quantum corrections
because there are large hierarchies
between the first and the third generations
and also
between the second and the third generations
on the analogy of Type A$^{(2)}$,

\noindent
\UL{case (b1)}: $diag(m_1^{},m^{}_2,0)$

$\sin^2 2\theta_{12}$ is unstable against quantum corrections.
 This is understood on the analogy of Type B1$^{(2)}$.
Thus, the MNS matrix is unstable against quantum corrections.

\noindent
\UL{case (b2)}: $diag(m_1^{},-m^{}_2,0)$

$\sin^2 2 \hat{\theta}_{12}$ is also stable against
quantum corrections analogous to the case of Type B2$^{(2)}$.
The MNS matrix is stable against quantum corrections.

\item
Type C ($m_1^{}\sim m_2^{}\sim m_3^{}$)

 The MNS matrix approaches to the definite unitary matrix according to the
relative sign assignments of neutrino mass eigenvalues,
as the effects of quantum corrections become
large enough to neglect squared mass differences of neutrinos.
 Independent parameters of the MNS matrix at the $M_R^{}$ scale
approach to the following fixed values in the large limit of 
quantum corrections.

\noindent
\UL{case (c1)}: $diag.(-m_1^{}, m_2^{},m_3^{})$ 
\beq
U_{e2}=\dfrac{\sin\theta}{\sqrt{1+\cos^2\theta}}\,,
\mbox{\quad}
U_{e3}=-\dfrac{1}{2}\dfrac{\sin 2\theta}{\sqrt{1+\cos^2\theta}}\,,
\mbox{\quad}
U_{\mu 3}=\dfrac{1}{\sqrt{2}}\dfrac{\sin^2\theta}{\sqrt{1+\cos^2\theta}}\,.
\eeq

\noindent
\UL{case (c2)}: $diag.( m_1^{},-m_2^{},m_3^{})$ 
\beq
U_{e2}=\sin \theta\,,
\mbox{\quad}
U_{e3}=\dfrac{1}{2}\dfrac{\sin 2 \theta}{\sqrt{1+\sin^2\theta}}\,,
\mbox{\quad}
U_{\mu 3}=\dfrac{1}{\sqrt{2}}\dfrac{\cos^2\theta}{\sqrt{1+\sin^2\theta}}\,.
\eeq

\noindent
\UL{case (c3)}: $diag.(-m_1^{},-m_2^{},m_3^{})$ 
\beq
U_{e2}=0\,,
\mbox{\quad}
U_{e3}=0\,,
\mbox{\quad}
U_{\mu 3}=\dfrac{1}{\sqrt{2}}\,.
\eeq

\noindent
\UL{case (c4)}: $diag.( m_1^{}, m_2^{},m_3^{})$ 
\beq
U_{e2}=0\,,
\mbox{\quad}
U_{e3}=0\,,
\mbox{\quad}
U_{\mu 3}=0\,.
\eeq

\bs
\end{enumerate}

 Results of this article are not only
useful for the model building,
but also show the possibility to obtain the large mixing angles
from quantum corrections.

\section*{Acknowledgments}
One of the author NO thanks G.-C.~Cho and K.~Hagiwara
for useful discussion and comments.
The work of NH is partially supported by DOE grant DOE/ER/01545-753. 
 The work of NO is supported by the JSPS
Research Fellowships for Young Scientists, No. 2996.

\newpage
\appendix
\section{Oscillation Probabilities in the Vacuum}
\label{app:MNS}
\setcounter{equation}{0}
The transition probability $P_{\nu_\alpha \to \nu_\beta}$,
 $(\alpha \neq \beta)$ 
and the survival probability $P_{\nu_\alpha \to \nu_\alpha}$ 
of the neutrino-oscillation 
are given by 
\bea
P_{\nu_\alpha \to \nu_\beta} &=&
\l|\sum_{j=1}^3 \l(V_{\rm MNS}^{}\r)_{\beta j}^{} 
\exp\l({\dfrac{-i m_j^2}{2E}L}\r)
\l(V_{\rm MNS}^{\dagger}\r)_{j \alpha}^{} \r|^2\,, \nn \\
 &=& 
\l| U_{\beta 1}^{} U_{\alpha 1}^{\ast}
+U_{\beta 2}^{}
  \exp\l({\dfrac{-i \delta m_{12}^2}{2E}L}\r)
  U_{\alpha 2}^{\ast}
+U_{\beta 3}^{}
  \exp\l({\dfrac{-i \delta m_{13}^2}{2E}L}\r)
  U_{\alpha 3}^{\ast}
\r|^2,
\eqlab{pro1} \\
\nn \\
P_{\nu_\alpha \to \nu_\alpha} &=&
\l|\sum_{j=1}^3 \l(V_{\rm MNS}^{}\r)_{\alpha j}^{}
 \exp\l({\dfrac{-i m_j^2}{2E}L}\r)
\l(V_{\rm MNS}^{\dagger}\r)_{j \alpha}^{}\r|^2 \,,\nn \\
 &=& 
\l|U_{\alpha 1}^{} U_{\alpha 1}^{\ast}
+U_{\alpha 2}^{}
  \exp\l({\dfrac{-i \delta m_{12}^2}{2E}L}\r)
  U_{\alpha 2}^{\ast}
+U_{\alpha 3}^{}
  \exp\l({\dfrac{-i \delta m_{13}^2}{2E}L}\r)
  U_{\alpha 3}^{\ast}
\r|^2\,,
\eqlab{app-surv}
\eea
respectively.
Here we note $\delta m_{ij}^2 = m_j^2 - m_i^2$.
When 
$\delta m_{12}^2 \ll \delta m_{13}^2\,,$
eqs.(\ref{eqn:pro1})and (\ref{eqn:app-surv})
can be simplified as follows.

{\sl \UL{Condition 1}:}
\beq
\dfrac{\delta m_{12}^2}{2E}L \ll 1
\sim \dfrac{\delta m_{13}^2}{2E}L
\eqlab{A-massd}
\eeq
Under this condition,
\eqref{pro1}
becomes
\bea
P_{\nu_\alpha \to \nu_\beta} &=&
\l|-U_{\beta 3}^{} U_{\alpha 3}^{\ast}
+U_{\beta 3}^{}\exp\l({\dfrac{-i \delta m_{13}^2}{2E}L}\r) U_{\alpha 3}^{\ast}
\r|^2\,, \nn \\
&=&
4\l|U_{\alpha 3}^{}\r|^2\l|U_{\beta 3}^{}\r|^2 \sin^2
\l(
{\dfrac{\delta m_{13}^2}{4E}L}
\r)\,,
\eqlab{pro2}
\eea
and \eqref{app-surv} becomes
\bea
P_{\nu_\alpha \to \nu_\alpha} &=&
\l|1-U_{\alpha 3}^{} U_{\alpha 3}^{\ast}
 +U_{\alpha 3}^{}
  \exp\l({\dfrac{-i \delta m_{13}^2}{2E}L}\r)
  U_{\alpha 3}^{\ast}
\r|^2\,, \nn \\
&=&
 1 - 4\l|U_{\alpha 3}^{}\r|^2 \l(1-\l|U_{\alpha 3}^{}\r|^2 \r)
\sin^2
\l(
{\dfrac{\delta m_{13}^2}{4E}L}
\r).
\eqlab{surv}
\eea

{\sl \UL{Condition 2}:}
\beq
\dfrac{\delta m_{12}^2}{2E}L \sim 1 
\ll \dfrac{\delta m_{13}^2}{2E}L
\label{ap-cond-2}
\eeq
Under this condition,
\eqsref{pro1} and \eqvref{app-surv}
become
\bea
P_{\nu_{\alpha} \to \nu_{\beta}}
&=&
2|U_{\alpha 3}^{}|^2|U_{\beta 3}^{}|^2
-\l[
4 Re(U_{\alpha 1}^{}U_{\beta 1}^{\ast}U_{\beta 2}^{}U_{\alpha 2}^{\ast})
\sin^2\l(\dfrac{\delta m_{12}^2}{4E}L\r)
+2 J_{\rm MNS}^{}
\sin \l(\dfrac{\delta m_{12}^2}{2E}L\r)
\r]\,,\nn \\
 \\
P_{\nu_\alpha \to \nu_\alpha} &=&
1
-2 |U_{\alpha 3}^{}|^2 \l(1-|U_{\alpha 3}^{}|^2\r)
-4 |U_{\alpha 1}^{}|^2 |U_{\alpha 2}^{}|^2 
\sin^2
\l(
{\dfrac{\delta m_{12}^2}{4E}L}
\r)\,,
\eqlab{A9}
\eea
respectively.
Here $J_{\rm MNS}^{}$ is defined by $Im(U_{\alpha 1}^{}U_{\beta 1}^{\ast}
 U_{\beta 2}^{} U_{\alpha 2}^{\ast}$).

\section{Quantum Corrections of $\kappa$}
\label{app:RGE-K}
\setcounter{equation}{0}
In this section, we show the relation between $\kappa(m_z^{})$
and $\kappa(M_R^{})$ in the cases of
$\kappa_{33}=0$ and
$\kappa_{11}=\kappa_{22}=\kappa_{33}=0$.

\subsection{$\kappa_{33}=0$}
 At first, we discuss the case of $\kappa_{33}=0$
in the diagonal base of $\he$.
 If some elements of $\kappa$ are zero at $m_z^{}$,
they are always zero through all energy-scale.
 This means if $\kappa_{33}$ is zero at $m_z^{}$,
$\kappa$ cannot be normalized by $\kappa_{33}$
and $c_{ij}^{}$ of \eqref{def_kappa} cannot be defined.
 Thus, we adopt $\kappa_{22}$ for the normalization of $\kappa$ as
\bea
\kappa &=&
\kappa_{22}^{}
\bmaT
r^{\prime}_{1} & c^{\prime}_{12} \sqrt{r^{\prime}_{1}} &
  c^{\prime}_{13} r^{\prime}_{2} \sqrt{r^{\prime}_{1}} \\
c^{\prime}_{12} \sqrt{r^{\prime}_{1}} & 1 & r^{\prime}_{2}\\
 c^{\prime}_{13} r^{\prime}_{2}\sqrt{r^{\prime}_{1}} & 
r^{\prime}_{2} & 0
\emaT
\nn\,, \\
 &=&
\kappa_{22}^{}
\bmaT
\sqrt{r^{\prime}_{1}} & 0 & 0 \\
0                     & 1 & 0 \\
0                     & 0 & r^{\prime}_{2}  \\
\emaT
\bmaT
1 & c^{\prime}_{12} &  c^{\prime}_{13} \\
c^{\prime}_{12} & 1 &  1\\
c^{\prime}_{13} & 1 & 0
\emaT
\bmaT
\sqrt{r^{\prime}_{1}} & 0 & 0 \\
0                     & 1 & 0 \\
0                     & 0 & r^{\prime}_{2}  \\
\emaT\,,
\eqlab{k33=0_def}
\eea
where
$c^{\prime}_{1j}$s ($j=2,3$) are defined as
\beq
\l(c^{\prime}_{12}\r)^2 = \dfrac{\kappa_{12}^2}{\kappa_{11}\kappa_{22}}\,,
\mbox{\quad and \quad}
\l(c^{\prime}_{13}\r)^2 = 
\dfrac{\kappa_{22} \kappa_{13}^2}{\kappa_{11}\kappa_{23}^2}\,,
\eeq
which are energy-scale independent complex parameters.
$r^{\prime}_{1,2}$ in \eqref{k33=0_def} are defined as
\beq
r^{\prime}_{1} = \dfrac{\kappa_{11}}{\kappa_{22}}\,,
\mbox{\quad and \quad}
r^{\prime}_{2} = \dfrac{\kappa_{23}}{\kappa_{22}}\,.
\eeq
By using the notation of \eqref{Ii}, we can obtain
the energy-scale dependences of $r_{1,2}^{\prime}$ as
\beq
r^{\prime}_{1}(M_R^{}) = r^{\prime}_{1}(m_z^{}) \dfrac{I_e}{I_\mu} \,,
\mbox{\quad and \quad}
r^{\prime}_{2}(M_R^{}) = r^{\prime}_{2}(m_z^{})
\sqrt{\dfrac{I_\tau}{I_\mu}}\,.
\eqlab{Ii_33=0}
\eeq
Then, $\kappa(M_R^{})$ is given by
\beq
\kappa(M_R^{})=
\dfrac{k_{22}^{}(M_R^{})}{k_{22}^{}(m_z^{})}
\dfrac{I_\tau}{I_\mu}
\bmaT
\sqrt{I_e/I_\tau} &          0         &   0 \\
0                 &\sqrt{I_\mu/I_\tau} &   0 \\
0                 &  0                 &   1 \\
\emaT
{\kappa(m_z)}
\bmaT
\sqrt{I_e/I_\tau} &          0         &   0 \\
0                 &\sqrt{I_\mu/I_\tau} &   0 \\
0                 &  0                 &   1 \\
\emaT\,.
\eqlab{def_kappa33=0}
\eeq

\subsection{$\kappa_{11}=\kappa_{22}=\kappa_{33}=0$}
When all diagonal elements are zero,
all off-diagonal elements of $\kappa$ can be taken real.
It is because all phases are absorbed by the field redefinitions of
\beq
L_i \rightarrow e^{-i\varphi_i} L_i\,,
\mbox{\quad and \quad}
E_i \rightarrow e^{i\varphi_i} E_i\,,
\eeq
where $\varphi_i$s are defined as
\bea
\varphi_1 &=&
 \l(arg.(\kappa_{12})+arg.(\kappa_{13})-arg.(\kappa_{23})\r)/4\,, \nn \\
\varphi_2 &=&
 \l(arg.(\kappa_{12})-arg.(\kappa_{13})+arg.(\kappa_{23})\r)/4\,, \nn \\
\varphi_3 &=&
\l(-arg.(\kappa_{12})+arg.(\kappa_{13})+arg.(\kappa_{23})\r)/4\,.
\eea
Normalizing all elements  by
$\kappa_{23}$, 
$\kappa$ is given by
\bea
\kappa&=& \kappa_{23}
\bmaT
     0              & r^{\prime\prime}_{1} & r^{\prime\prime}_{2} \\
r^{\prime\prime}_{1}&      0               & 1 \\
r^{\prime\prime}_{2}& 1                    & 0 \\
\emaT\,, \nn \\
&=&\kappa_{23}
\bmaT
\sqrt{r^{\prime\prime}_{1}r^{\prime\prime}_{2}} & 0 & 0 \\
0 &\sqrt{r^{\prime\prime}_{1}/r^{\prime\prime}_{2}}  & 0 \\
0 & 0 &\sqrt{r^{\prime\prime}_{2}/r^{\prime\prime}_{1}}
\emaT
\bmaT
0 & 1 & 1 \\
1 & 0 & 1 \\
1 & 1 & 0
\emaT
\bmaT
\sqrt{r^{\prime\prime}_{1}r^{\prime\prime}_{2}} & 0 & 0 \\
0 &\sqrt{r^{\prime\prime}_{1}/r^{\prime\prime}_{2}}  & 0 \\
0 & 0 &\sqrt{r^{\prime\prime}_{2}/r^{\prime\prime}_{1}}
\emaT\,,
\nn \\
\eqlab{Zee-kappa2}
\eea
where
\beq
r^{\prime\prime}_{1} = \dfrac{\kappa_{12}}{\kappa_{23}}\,,
\mbox{\quad and \quad}
r^{\prime\prime}_{2} = \dfrac{\kappa_{13}}{\kappa_{23}}\,.
\eeq
By using the notation of \eqref{Ii}, we can obtain
$r^{\prime\prime}_{1,2}(M_R^{})$ as
\beq
r^{\prime\prime}_{1}(M_R^{}) = 
r^{\prime\prime}_{1}(m_z^{})\sqrt{\dfrac{I_e}{I_\tau}}\,,
\mbox{\quad and \quad}
r^{\prime\prime}_{2}(M_R^{}) =
r^{\prime\prime}_{2}(m_z^{})\sqrt{\dfrac{I_e}{I_\mu }}\,.
\eeq
Then, $\kappa(M_R^{})$ is given by
\beq
\kappa(M_R^{})=
\dfrac{\kappa_{23}^{}(M_R^{})}{\kappa_{23}^{}(m_z^{})}
\sqrt{\dfrac{I_\tau}{I_\mu}}
\bmaT
\sqrt{I_e/I_\tau} &          0         &   0 \\
0                 &\sqrt{I_\mu/I_\tau} &   0 \\
0                 &  0                 &   1 \\
\emaT
{\kappa(m_z)}
\bmaT
\sqrt{I_e/I_\tau} &          0         &   0 \\
0                 &\sqrt{I_\mu/I_\tau} &   0 \\
0                 &  0                 &   1 \\
\emaT\,.
\eqlab{def_zee}
\eeq

Equations (\ref{eqn:def_kappa33=0}) and (\ref{eqn:def_zee})
show that the energy-scale dependence of the MNS matrix
can be estimated by $\sqrt{I_e/I_\tau}$
and $\sqrt{I_\mu/I_\tau}$
even in the cases of $\kappa_{33}=0$ or 
$\kappa_{11}=\kappa_{22}=\kappa_{33}=0$.
 In general, quantum corrections
of the MNS matrix can be estimated by only $n_g^{}-1$
degrees of freedom.
 It is easily understood as follows.
 The RGE of $\kappa$ (\eqref{RGE_kappa})
is separated into two parts,
which are
the lepton-flavor independent terms
(the gauge and the Higgs particles corrections)
and the lepton-flavor dependent terms
(the charged-lepton corrections).
 The energy-scale dependences of the mixing angles of $\kappa$
are determined by the flavor-dependent corrections
from $\he$.
Since $\he$ has $n_g^{}$ degrees of freedom,
expect for the over all factor,
$n_g^{}-1$ degrees of freedom decide the 
energy-scale dependence of the MNS matrix
as are shown in \eqsref{def_kappa2}, \eqvref{def_kappa33=0}
and \eqvref{def_zee}.

\section{Approximation of the Renormalization Corrections}
\label{app:Itau}
\setcounter{equation}{0}

We show the approximation of $\sqrt{I_i/I_\tau}$.
 Although we do not use this approximation in our numerical analyses,
it is useful for the rough estimation of quantum corrections of $\kappa$.

 At first, let us estimate the values of 
\beq
\dfrac{\sqrt{I_{e,\mu}/I_{\tau}} - \sqrt{1/I_\tau}}
      {\sqrt{I_{e,\mu}/I_{\tau}}}
=1- \dfrac{1}{\sqrt{I_{e,\mu}}}\,.
\eqlab{chIt1}
\eeq
 The magnitudes of \eqref{chIt1} are estimated to be
\beq
0<
1- \dfrac{1}{\sqrt{I_e}}
\ll 10^{-8}\,,
\mbox{\quad and \quad}
0<
1- \dfrac{1}{\sqrt{I_\mu}}
\ll 10^{-3}\,,
\eqlab{It-check}
\eeq
in the region of $2\leq\tan\beta \leq 60$ from the numerical analysis
shown in \Fgref{checkIt}.
\begin{figure}[htb]
\begin{center}
 {\scalebox{0.62}{\includegraphics{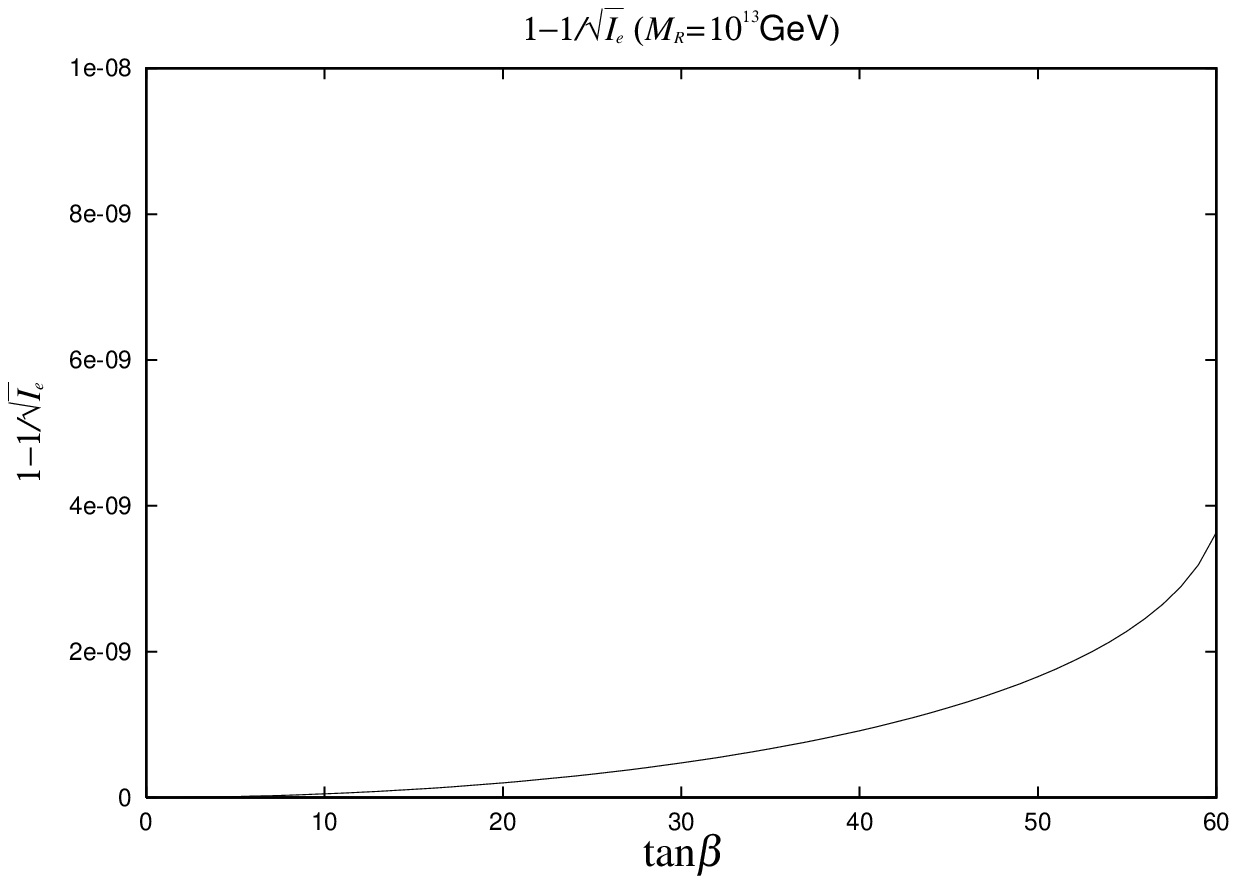}}  } 
 {\scalebox{0.62}{\includegraphics{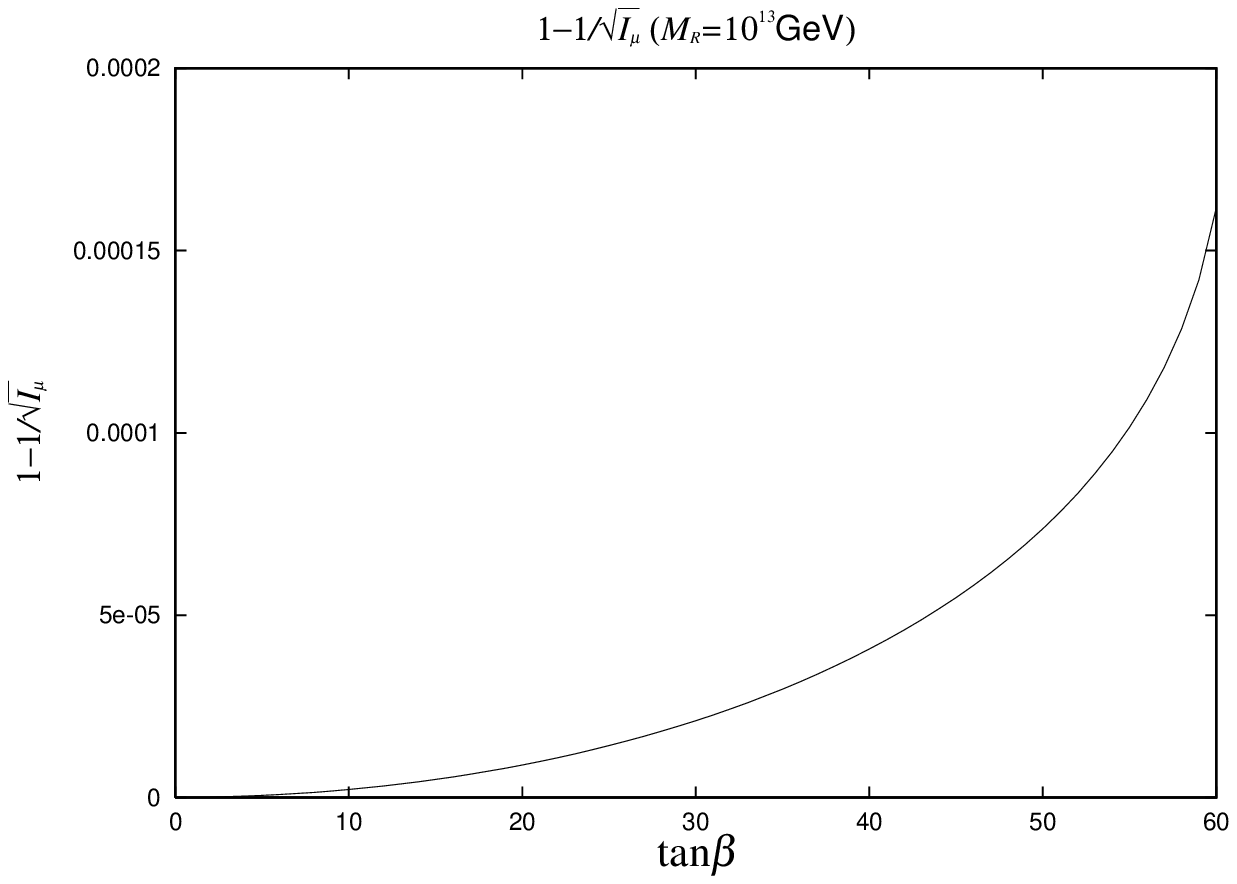}}  } 
\end{center}
 \vspace{-1.5em}
\caption{$\tan \beta$ dependences of \eqsref{chIt1}.} 
 \fglab{checkIt}
\end{figure}
 Thus, the approximation of
\beq
\sqrt{\dfrac{I_{e,\mu}}{I_\tau}} \simeq
\dfrac{1}{\sqrt{I_\tau}}
\eqlab{It-check2}
\eeq
is held with good accuracy.
 If we neglect the energy-scale dependence of $y_\tau$,
the value of \eqref{It-check2} is given by
\bea
\ln\l(\dfrac{1}{\sqrt{I_\tau}}\r)
&=&{\dfrac{1}{8\pi^2}\l(\dfrac{m_\tau}{v}\r)^2\l(\tan \beta^2 +1 \r) }
\ln\l(\dfrac{m_z^{}}{M_R^{}}\r)\,
\eqlab{check1}
\eea 
from \eqref{Ii},
where $m_\tau$ is the mass of $\tau$-lepton and 
$v^2=\vev{\phi_{\Sd}^2}+\vev{\phi_{\Su}^2}$.
 We define the faction of
\beq
{Err}(\tan \beta,M_R^{})=
1-{\sqrt{I_\tau}}\times\l(\dfrac{m_z^{}}{M_R^{}}\r)^
{\dfrac{1}{8\pi^2}\l(\dfrac{m_\tau}{v}\r)^2\l(\tan \beta^2 +1 \r) }
\eqlab{check11}
\eeq
to check the accuracy of \eqref{check1}.
 The $\tan \beta$ dependence of $|Err|$
is shown in \Fgref{check1} with $M_R^{}=10^{13}$ GeV.
\begin{figure}[htb]
\begin{center}
 {\scalebox{1.}{\includegraphics{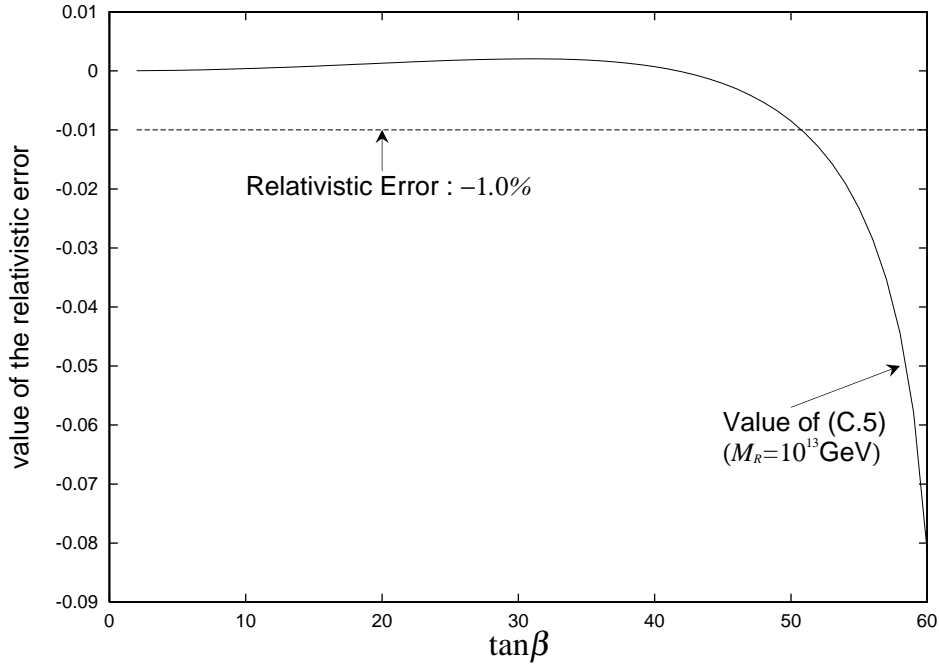}}  } 
\end{center}
 \vspace{-1.5em}
\caption{$\tan \beta$ dependence of \eqref{check11}.} 
 \fglab{check1}
\end{figure}
In the region of $2\leq\tan \beta \leq 50$,
the error of \eqref{check1} is less than 1\%.
Even in the region of $50 \leq \tan \beta \leq 60$, 
where the energy scale dependence of $y_\tau$ cannot be neglected,
$|Err|$ is less than 10\%.

 If $M_R^{}$ is smaller than $10^{13}$ GeV, 
the approximation of \eqref{check1}
becomes more accurate
because the Yukawa couplings of the charged-leptons are
enhanced in the high energy scale.

\end{document}